\documentclass[a4paper]{article}
\pdfoutput=1 
\usepackage[a4paper]{geometry}
\usepackage{hyperref}

\usepackage{graphicx}
\usepackage[intlimits]{amsmath}
\usepackage{amsfonts}
\usepackage{url}
\usepackage{theorem}
\newtheorem{problem}{Problem}
\newtheorem{definition}{Definition}
\title{Faster Computation of Expected Hypervolume Improvement}
\author{Iris Hupkens \and
        Michael Emmerich \and Andr{\'e} Deutz}

\date{Leiden Institute for Advanced Computer Science, \\ Leiden University, Niels Bohrweg 1, 2333-CA Leiden \\
              Tel.: +31-71-527-7094\\
              Fax: +31-71-527-7001\\
              emmerich AT liacs.nl   
}

\begin{document}
\large
\maketitle

\begin{abstract}
The expected improvement algorithm (or efficient global optimization) aims for global continuous optimization with a limited budget of black-box function evaluations. It is based on a statistical model of the function learned from previous evaluations and an infill criterion - the expected improvement - used to find a promising point for a new evaluation. The `expected improvement' infill criterion takes into account the mean and variance of a predictive multivariate Gaussian distribution.

The expected improvement algorithm has recently been generalized to multiobjective optimization. In order to measure the improvement of a Pareto front quantitatively the gain in dominated (hyper-)volume is used. The computation of the expected hypervolume improvement (EHVI) is a multidimensional integration of a step-wise defined non-linear function related to the Gaussian probability density function over an intersection of boxes. This paper provides a new algorithm for the exact computation of the expected improvement to more than two objective functions. For the bicriteria case it has a time complexity in $O(n^2)$ with $n$ denoting the number of points in the current best Pareto front approximation. It improves previously known algorithms with time complexity $O(n^3 \log n)$. For tricriteria optimization we devise an algorithm with time complexity of $O(n^3)$.  Besides discussing the new time complexity bounds the speed of the new algorithm is also tested empirically on test data. It is shown that further improvements in speed can be achieved by reusing data structures built up in previous iterations. The resulting numerical algorithms can be readily used in existing implementations of hypervolume-based expected improvement algorithms.
\bigskip

\end{abstract}


\newpage
\section{Introduction}


In multiobjective optimization, the goal is to find (a set of) solutions which optimize multiple objective functions at the same time \cite{miettinen}. As in case of conflicting objectives there is no single best solution, it is common to compute instead all points of the Pareto front (PF) or an approximation to this set.

Sometimes the function values of solutions can only be determined through costly simulations, so approximation functions are used in their place.
This makes it possible to evaluate the function values of the most promising individuals only, instead of wasting time evaluating the function values of individuals that are unlikely to result in an improvement.  A common approximation method are Gaussian processes (or Kriging), which yield a prediction in the form of a 1-D normal distribution. This {\em predictive distribution} is learned from previously evaluated points and quantifies for a given new point how likely it is that certain function values will be obtained. In the context of computer experiments, the statistical assumptions of such metamodels were discussed in Sacks et al. \cite{sachs1989}.

Optimization methods for expensive function values based on Gaussian processes date back to the Lithuanian school of global optimization \cite{mockusei}. More recently they have been refined and gained popularity in the context of optimization with expensive computer experiments \cite{jonesei} under names such as {\em efficient global optimization} (EGO) and {\em expected improvement algorithm} \cite{eiconvergence}.
Even more recently, different expected improvement formulations for multiobjective optimization have been developed and were compared in Wagner et al \cite{hypermonotonicity}. Among these, the {\em expected hypervolume improvement} (EHVI) turns out to have desirable theoretical properties.  The EHVI was first suggested in \cite{emmerichphd} and represents the expected improvement in the hypervolume measure relative to the current approximation of the Pareto front \cite{hypermonotonicity} given the probability distribution of possible function values. The hypervolume measure itself is a common measure used to determine the quality of a set of solutions to a multiobjective optimization problem \cite{hypervolume} and can be applied without a priori knowledge of the Pareto front, which makes the EHVI a natural quality measure to use in multiobjective surrogate-assisted optimization.

The calculation of the EHVI has so far been a problem. Monte Carlo integration can solve the issue of computing the EHVI directly, but to get an accurate approximation out of Monte Carlo integration is slow. An exact calculation approach exists for the bi-objective case, but it is slow as well (time complexity in $O(n^3 \log n)$). This thesis aims to increase the speed of the exact calculation of the EHVI in two dimensions, as well as provide a method of calculating it in higher dimensions.  Its implementation will be validated with results from Monte Carlo integration. The empirical performance of directly calculating the EHVI in the three-dimensional case will also be analyzed in order to show the feasibility of using direct calculations in place of Monte Carlo integration. The main contribution of this paper is therefore to make the EHVI computation both exact and fast, so that it can be used in Gaussian-process assisted global optimization algorithms.

The article is structured as follows: Section \ref{preliminaries} introduces important definitions and technical preliminaries. Section \ref{relatedwork} summarizes the related work.
Section \ref{2dcomplexity} contains a proof that the exact calculation of the bi-objective EHVI can be done in $O(n^2)$ as well as a lower bound on the worst case complexity of $\Omega(n \log n)$.  The proof of the upper bound is by constructing an algorithm. The implementation of this algorithm is then empirically compared to the naive ($O(n^3 \log n)$) implementation.
Section \ref{higherdimensionalehvi} describes the details of the new, general algorithm for calculating the expected hypervolume improvement in more than two dimensions, and Section \ref{sliceupdates} describes an exact method for determining the tri-objective EHVI with time complexity $O(n^3)$.
Section \ref{emptests} describes the results of empirical tests of implementations of the algorithms described in Sections \ref{higherdimensionalehvi} and \ref{sliceupdates}, both to validate the correctness of the implementations and to measure their performance. Finally, Section \ref{thefuture} contains concluding remarks and an outline of promising directions for future research.

\section{Preliminaries}
\label{preliminaries}
Without loss of generality, we will consider maximization of $m\geq1$ objective functions $f_1: X \rightarrow \mathbb{R}$, $\dots,$ $f_m: X \rightarrow \mathbb{R}$. A distinction is made between the decision space $X$ of alternative solutions and the objective space $\mathbb{R}^m$ where the images of points in $X$ under $\mathbf{f}$ are represented. However, the attention of this study will be on points and probability distributions of points in the objective space.

The {\em a posteriori} approach of multiobjective optimization is concerned with finding (approximations to) the Pareto front, that is: the set of solutions in the objective space that are not dominated in the Pareto dominance relation \cite{miettinen}.

The hypervolume indicator is a quality measure for  Pareto approximation sets \cite{zitzler}. Among the performance measures being used in Pareto optimization, it has some favorable properties. First of all it can be used to compute both absolute and relative improvements of a Pareto front approximation without a priori knowledge of the Pareto front and it is compliant with Pareto dominance \cite{viviane}. Furthermore, its maximization yields a set of Pareto optimal points distributed across the Pareto front \cite{fleischer,auger,bringmann}.

The {\em hypervolume indicator} of a finite set of points $P \subset \mathbb{R}^m$ with respect to a user-defined reference point $r$  is defined as the Lebesgue measure of the hypervolume covered by the boxes that have an element of $P$ as their upper corner and a reference point $r$ as their lower corner. Thereby it measures the size of the dominated space of $P$ cut from below by a reference point.  The reference point must be chosen in a such a way that it is dominated by all points in $P$, and ideally also by all points of the Pareto front.

%

The set containing the part of the objective space that is dominated by the points in $P$ will be referred to as $\mbox{DomSet}(P)$. The hypervolume contribution of a point $p \in P$ is the difference in dominated hypervolume between $P \setminus \{p\}$ and $P$. The hypervolume contribution of a set of points $S \subseteq P$ is defined analogously, as the difference between $P \setminus S$ and $P$.

The {\em hypervolume improvement} of a point $p \notin P$ with respect to $P$ is defined as the hypervolume contribution of $p$ with regards to $P \cup \{p\}$, i.\,e. the increment of the hypervolume indicator after $p$ is added to $P$.

Note that in the entire article we consider a fixed reference point that is dominated by all points in the Pareto front approximations. The choice of reference point is an important issue by itself, which we do however not adress here.

\subsection{The Expected Hypervolume Improvement}

In global optimization with expensive function evaluations it is common to predict function values using statistical methods such as Gaussian processes \cite{jonesei}. Such methods provide a predictive distribution of possible outcomes of the precise evaluation of the vector valued objective function in form of the parameters of a probability density function (PDF) over all possible outcomes. In case of Gaussian processes or Kriging approximations the predictions are given by multivariate normal distributions\cite{emmerichphd}.

The expected hypervolume improvement (EHVI) is the expected value of the hypervolume improvement of a new candidate point in $X$, given its predictive distribution function ($PDF$) over points in the objective space. The general formula for the EHVI with respect to a mutually non-dominated set $P$ is\cite{emmerichphd}:
\begin{equation}
\int_{p \in \mathbb{R}^m} \mbox{HI}(p,P) \cdot PDF(p)\mathrm{d}p
\end{equation}
The EHVI is a generalization of the classical expected improvement (EI) criterion $\int_{y=f'}^\infty \max\{0, y-f'\} PDF(y) dy$ used in model-assisted single objective optimization, where $f'$ denote the function value of the currently best solution\cite{mockusei,jonesei}.

For a given mean and standard deviation vector of an independently distributed predictive distribution, the EHVI is monotonic with respect to the mean value \cite{hypermonotonicity} and, at least for $m \leq 2$, also w.r.t. the variance\cite{exacthvi}. It has been used as an infill criterion for multiobjective EGO in multiobjective optimization in various studies \cite{zaefferer,Shimoyama12,Shimoyama13} but its application so far has been confined to the bi-objective case and the computation of the EHVI was criticized to be computationally expensive as compared to more simple generalizations of the EI\cite{hypermonotonicity}.

In \cite{exacthvi}, a formula is derived for exactly calculating the EHVI for $m=2$ independent and identically distributed normal PDFs. The expression in \cite{exacthvi} in general does not yield the exact result for $m>2$, as will be shown later.



%
%

Given these preliminaries the general problem discussed in this article can now be defined concisely:
\begin{problem}
Given a finite set of points  $P \subset \mathbb{R}^m$, a reference point $r$ and a predictive independent distributed multivariate normal PDF, given by its mean value $\mu \in \mathbb{R}^m$ and standard deviations $\sigma \in \mathbb{R}^m$, how can the EHVI be computed and how can the EHVI be computed efficiently?
\label{problem}
\end{problem}

\subsection{One-Dimensional Expected Improvement and its Decomposition}
\label{psiscription}

In order to calculate the EHVI, we will need to calculate many integrals that have the form of a partial one-dimensional improvement. In \cite{exacthvi}, a function was derived that could be used for that purpose.

In the following definition we recall the notion of standard normal distribution and normal distribution. Moreover we introduce a useful shorthand named $\psi$.

\begin{definition}
\begin{enumerate}
\item The function $\phi(s)= 1/\sqrt{2\pi}e^{-\frac{1}{2}s^2}, s\in \mathbb{R}$ is the density function of the standard normal distribution and $\Phi(s)= \frac{1}{2}\left(1 + \mbox{\mbox{erf}}\left(\frac{s}{\sqrt{2}}\right)\right)$ is the cumulative probability distribution function of the standard normal distribution. The general normal distribution with mean $\mu$ and variance $\sigma$ has as density the function $\phi_{\mu,\sigma}(s)= \frac{1}{\sigma\sqrt{2\pi}}e^{-\frac{1}{2}(\frac{s-\mu}{\sigma})^2}, s\in \mathbb{R}$. The cumulative distribution function of the general normal distribution is:  $\Phi_{\mu, \sigma}(s)= \frac{1}{2}\left(1 + \mbox{\mbox{erf}}\left(\frac{s-\mu}{\sigma \sqrt{2}}\right)\right)$.
\item \begin{equation}\psi(a,b,\mu,\sigma) :
=  \sigma \cdot \phi(\frac{b - \mu}{\sigma}) + (a - \mu)\Phi(\frac{b - \mu}{\sigma}) \end{equation}
\end{enumerate}
\end{definition}
Remark: it is easy to check that $\phi_{\mu, \sigma}(s) = \frac{1}{\sigma}\phi(\frac{s-\mu}{\sigma}) $ and $\Phi_{\mu, \sigma}(s) = \Phi(\frac{s-\mu}{\sigma})$. \\



Integrals of the form $\int_{z = b}^\infty (z - a) \frac{1}{\sigma}\phi(\frac{z - \mu}{\sigma})$ are equal to $\sigma\phi(\frac{b-\mu}{\sigma}+(\mu-a)[1-\Phi(\frac{b-\mu}{\sigma})]$.
Integrals whose upper limit is less than $\infty$ and lower limit greater than $-\infty $ can be written as the difference of two such integrals, allowing partial expected improvements over an interval $[l,u) \subset \mathbb{R}$, $l \geq f'$ to be calculated. Moreover one can easily see that this difference can be neatly expressed in terms of $\psi$:

\begin{equation*}
\begin{split}&\int_{z = l}^u (z - f')\frac{1}{\sigma}\phi(\frac{z - \mu}{\sigma})\mathrm{d}z \\
=& \int_{z = l}^\infty (z - f')\frac{1}{\sigma}\phi(\frac{z - \mu}{\sigma})\mathrm{d}z - \int_{z = u}^\infty (z - f')\frac{1}{\sigma}\phi(\frac{z - \mu}{\sigma})\mathrm{d}z \\
=& \: \psi(f',l,\mu,\sigma) - \psi(f',u,\mu,\sigma)
\end{split}
\end{equation*}

The value $f'$ in this case is the currently best function value.

In the rest of this thesis we will use the abbreviations $\phi_x(s) := \phi_{\mu_x, \sigma_x}(s) (= \frac{1}{\sigma_x}\phi(\frac{s - \mu_x}{\sigma_x}) )$ and $\Phi_x(s) := \Phi_{}(s) (= \Phi(\frac{s - \mu_x}{\sigma_x}) )$, where $\mu_x$ and $\sigma_x$ are the mean and variance of the normal distribution associated to the point $x$ in the search space.  Analogously, we use abbreviations $\phi_y, \Phi_y$ and $\phi_z, \Phi_z$ for the $y$ and the $z$ coordinate.


\section{Related Work}
\label{relatedwork}
The use of the one-dimensional expected improvement to solve engineering problems with expensive-to-evaluate objective functions was initially proposed by Mockus et al.\cite{mockusei} and then later reintroduced by Jones et al. in \cite{jonesei}. Since then it  has been widely used in global optimization with expensive-to-evaluate functions. It has been shown to converge to the global optimum for the single objective case and a subclass of continuous functions \cite{eiconvergence}.


%
Generalizing the one-dimensional expected improvement algorithm to multiobjective optimization is still a very new area of research. Besides the aforementioned EHVI, first published in \cite{emmerichphd}, various other solutions have been proposed:
\begin{itemize}
\item Chebyshev scalarization with dynamically changing weights \cite{chebyscale}.
\item Scalarization by using the distance from the centroid of the probability distribution to the Pareto approximation set \cite{keaneei}.
\item The hypervolume improvement for candidate points, calculated based on the upper confidence bound of the meta-model prediction \cite{smsego}.
\end{itemize}
Moreover, in Kumano et al. \cite{kumano} it was proposed to use a vector of expected improvements for the single objective functions, instead of a scalar measure.

Expected improvement has been studied as an infill criterion in different application fields, such as bioinformatics \cite{chebyscale}, mechanical engineering \cite{smsego} and aerospace design \cite{keaneei}. Also the EHVI was already applied in practice for the tuning of controllers in sewage treatment plants \cite{zaefferer}, in mechanical engineering \cite{Shimoyama12} and quantum control \cite{Shir07}. As compared to other indicators EHVI was found to have monotonicity in mean values \cite{hypermonotonicity} and variance\cite{exacthvi} and yielded high accuracy optima approximations. However, its computation is so far limited to the biobjective case and the time complexity of existing exact algorithms is still very high ($O(n^3 \log n)$, see\cite{exacthvi}.

Recently, Couckuyt et al. \cite{Couckuyt13} published an algorithm that is faster based on empirical tests. The complexity of this algorithm is not reported. It follows a heuristic block partitioning scheme for computing the improvement contribution of each cell and we conjecture its total complexity to be in $\Omega(n^3)$ for two and in $\Omega(n^4)$ in three objectives.


\section{Calculating the 2-D Expected Hypervolume Improvement}
\label{2dcomplexity}
Firstly, an efficient exact algorithm for the computation of the EHVI in two dimensions will be discussed.

Let $P$ denote a set of $n$ mutually non-dominated points in the two-dimensional plane. $P$ is the currently best Pareto front approximation. Furthermore, let $r \in \mathbb{R}^2$ denote a reference point which is dominated by every point in $P$. The aim is to calculate the expected hypervolume improvement for a point $p$ in the decision space for which we have the mean $(\mu_x, \mu_y)$ and standard deviation $(\sigma_x, \sigma_y)$ of a predictive distribution.

\begin{figure}[!h]
\begin{center}
\includegraphics[width=60mm]{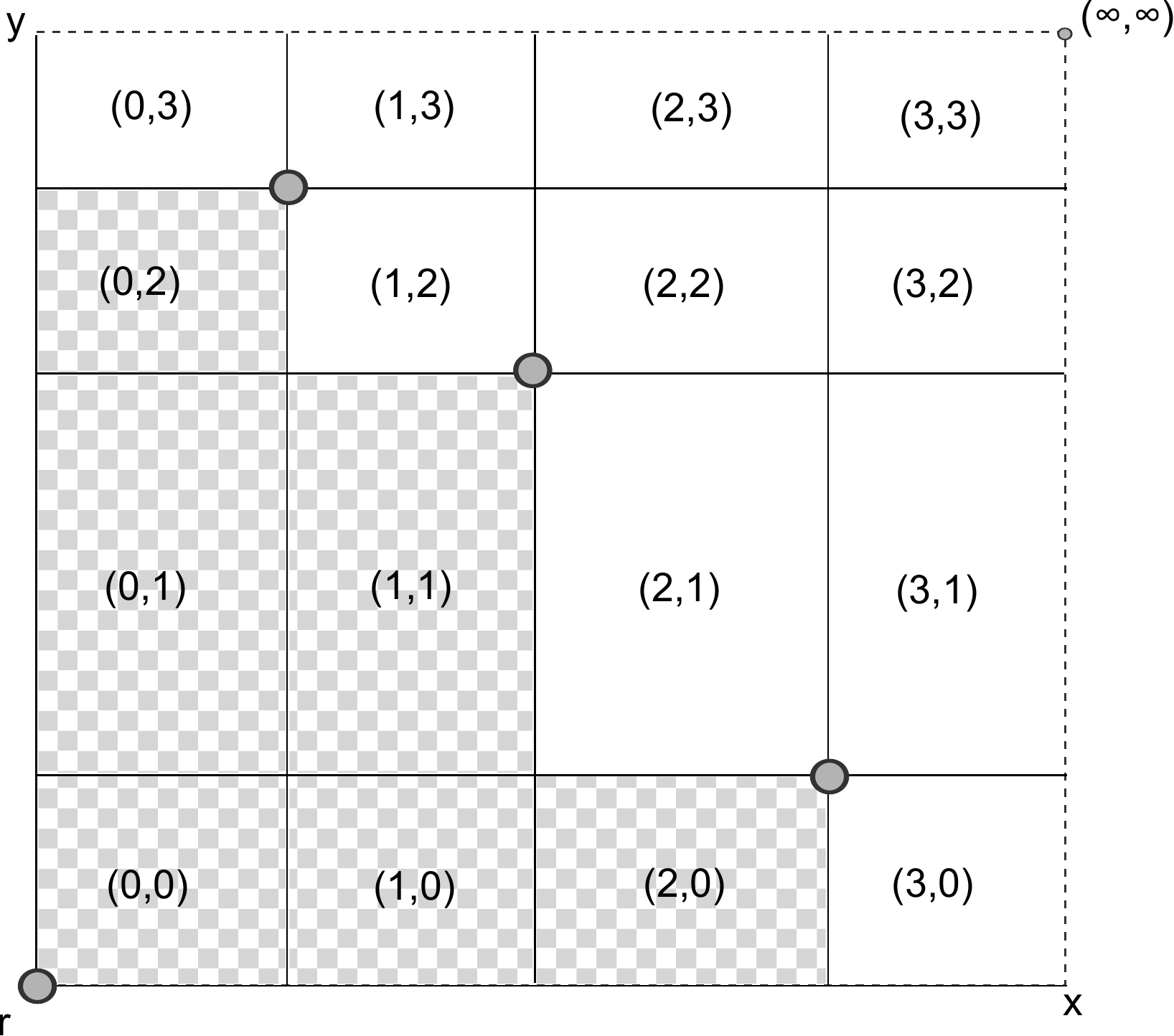}
\caption{An example of the interval boxes for a small population $P$. Checkered boxes fall in the dominated hypervolume of $P$. Therefore their contribution to the integral will be 0, and no calculation will be necessary for these boxes.}
\label{gridpicture}
\end{center}
\end{figure}

In the two-dimensional case, calculating the EHVI for $p$ exactly can be done by piecewise integration over a set of half-open rectangular interval boxes (cells) formed by the horizontal and vertical lines going through the points in $P$ and through $r$. The final EHVI is the sum of the contributions calculated for all grid cells. See Figure \ref{gridpicture} for a visualization of the grid.


Individual grid cells will be denoted by $C(a,b)$, where $0 \leq a \leq n$ and $0 \leq b \leq n$.  Let $Q = P \cup \{(\infty,r_y)\} \cup \{(r_x, \infty)\}$, with $Q^x$ denoting $Q$ sorted in order of ascending $x$ coordinate, and $Q^y$ denoting $Q$ sorted in order of ascending $y$ coordinate. Let $C$ be the set of grid cells representing the interval boxes. The numbers $a$ and $b$ represent positions in the sorting order of $Q$, starting with 0. Then, $a$ is the position of elements of $Q^x$ and $b$ is the position of elements of $Q^y$. The lower left corner of a cell will have the coordinates $(Q^x_a.x, Q^y_b.y)$. The upper right corner of the grid cell will have the coordinates $(Q^x_{a+1}.x,Q^y_{b+1}.y)$.


Note, that due to the characteristics of mutually non-dominated points in the two-dimensional plane, it is not necessary to sort $Q$ twice in order to determine $Q^x$ and $Q^y$. Sorting $P$ in order of ascending $x$ coordinate is equivalent to sorting it in order of descending $y$ coordinate. It follows that $Q^x_k = Q^y_{n+1-k}$.

When dividing the grid in the way described above, $(n+1)^2$ interval boxes are formed. However, if the upper right corner of an interval box is dominated by or equal to some point in $P$, its contribution will be zero, and no calculation will need to be done for that interval box.
Note that if the upper right corner is not dominated by $P$ then the lower left corner is neither.
These interval boxes are represented by a grid cell $C(a,b)$ which is within the dominated hypervolume of $P$. The remaining cells, $C_{stairs}$, are formed by cells for which this is not the case, meaning that $\forall (C(a,b) \in C_{stairs}, p \in P): p.x > Q^x_a.x \Rightarrow Q^y_b.y \geq p.y$ and, analogously, $p.y > Q^y_b.y \Rightarrow Q^x_a.x \geq p.x$.

Due to the definition of $Q$, we know that for $p \in P$ it holds that $p = Q^x_k = Q^y_{n+1-k}$ for some $0 < k \leq n$. Furthermore $k > a$ and $n+1-k > b$, if and only if $p$ dominates $C(a,b)$.
From this we get the following equivalence:
$a \geq n-b \Leftrightarrow C(a,b)$ is dominated by some point $p \in P$. Thus $C_{\mbox{stairs}}$ consists of all cells satisfying $a \geq n-b$.
There are $\frac{(n+1)(n+2)}{2}$ of such cells, resulting in a lower bound of $O(n^2)$ on the complexity of any algorithm which iterates over these interval boxes. 

If we call the lower corner of the cell $l$ and the upper corner $u$, the contribution of a grid cell to the integral is defined as follows:

$$\int_{p_y = l_y}^{u_y} \int_{p_x = l_x}^{u_x} \mbox{HI}(p) \,\phi_x(p_x) \, \phi_y(p_y) \, \mathrm{d}p_x\,\mathrm{d}p_y$$

\begin{figure}[!h]
\begin{center}
\includegraphics[width=60mm]{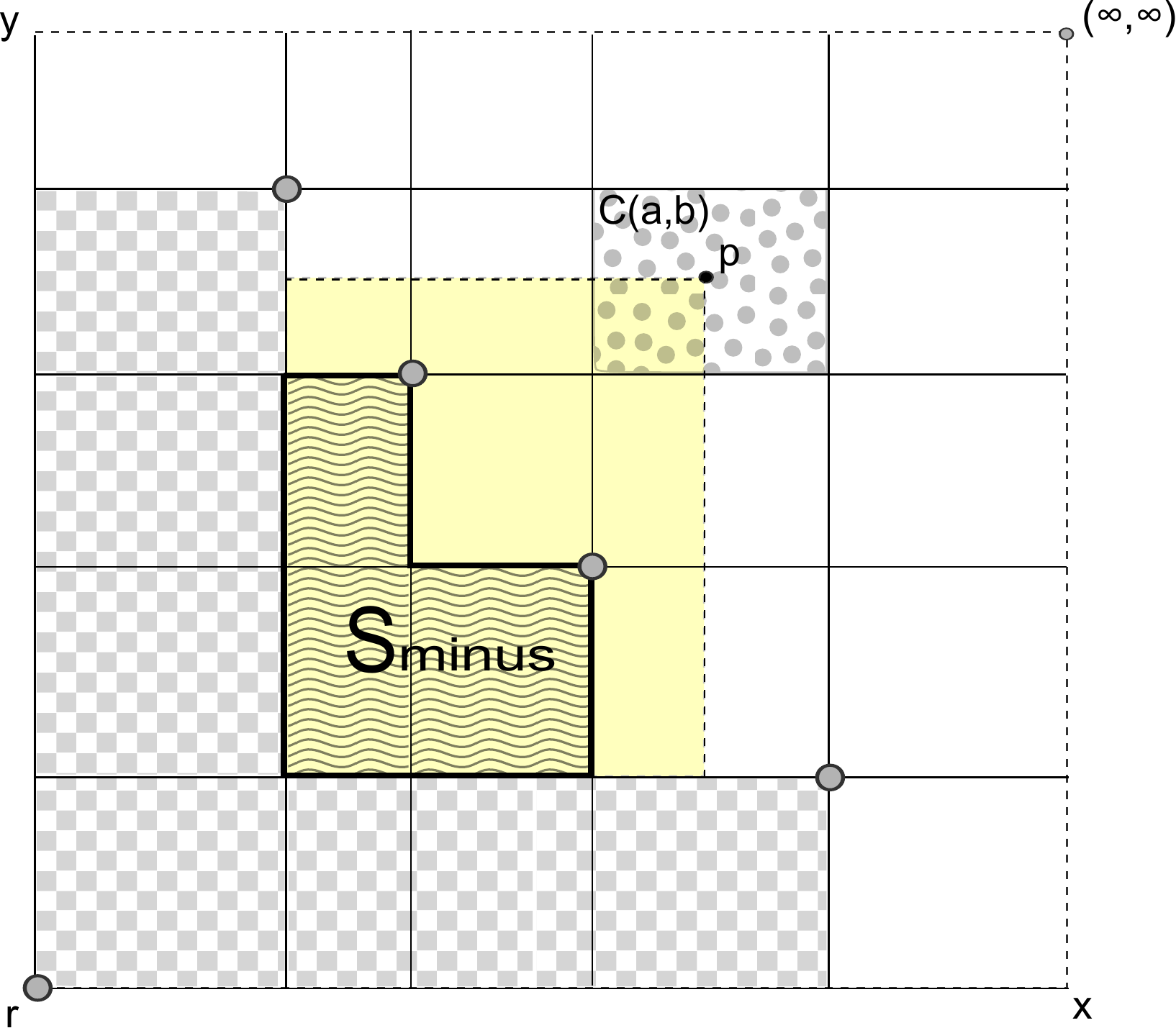}
\caption{Within an integration region $C(a,b)$, the hypervolume improvement of candidate points $p$ is equal to $(p.x - Q^y_{b+1}.x) \cdot (p.y - Q^x_{a+1}.y) - S_{minus}$. In this example, the yellow rectangle represents $(p.x - Q^y_{b+1}.x) \cdot (p.y - Q^x_{a+1}.y)$, and $S$ consists of the two points within the yellow rectangle.}
\label{Sminusbox}
\end{center}
\end{figure}

Dominated cells have a contribution of 0 to the integral, and for cells which are non-dominated, $\mbox{HI}(p)$ can be calculated as a rectangular volume from which a correction term is subtracted. See Figure \ref{Sminusbox} for a visual representation. The integral for these cells can be calculated as follows, as was described in more detail in \cite{exacthvi}:

\begin{equation*}
\begin{split}
& \int_{p_y = l_y}^{u_y} \int_{p_x = l_x}^{u_x} (p_x - r_x) (p_y - r_y) - S_{minus} \,\phi_x(p_x) \, \phi_y(p_y) \, \mathrm{d}p_x\,\mathrm{d}p_y \\
= &\int_{p_y = l_y}^{u_y} \int_{p_x = l_x}^{u_x} (p_x - v_x) (p_y - v_y) \,\phi_x(p_x) \, \phi_y(p_y) \, \mathrm{d}p_x\,\mathrm{d}p_y\\
& - \int_{p_y = l_y}^{u_y} \int_{p_x = l_x}^{u_x} S_{minus} \,\phi_x(p_x) \, \phi_y(p_y) \, \mathrm{d}p_x\,\mathrm{d}p_y \\
= & \, \left(\psi(v_x, l_x,\mu_x,\sigma_x)-\psi(v_x, u_x,\mu_x,\sigma_x)\right) \cdot \left(\psi(v_y, l_y,\mu_y,\sigma_y)-\psi(v_y, u_y,\mu_y,\sigma_y) \right)\\
&  - S_{minus} \cdot (\Phi_x(u_x)-\Phi_x(l_x) )\cdot( \Phi_y(u_y)-\Phi_y(l_y))
\end{split}
\end{equation*}

The last step is motivated by Subsection \ref{psiscription} and the application of Fubini's Theorem \cite{fubini}. It can be seen that the formula is of the form $c_1 - S_{minus} \cdot c_2$, where $c_1$ and $c_2$ are calculations which are performed in constant time with respect to $n$ for a single cell.

The correction term $S_{minus}$ is equal to the hypervolume contribution of $S \subseteq P$, where $S$ consists of those points dominated by or equal to the lower corner of the cell. Calculating the dominated hypervolume of a set in the two-dimensional plane has a time complexity of $O(n \log n)$. This complexity results from needing to find the neighbors of each point in order to calculate its contribution to the hypervolume. Sorting the set has a time complexity of $O(n \log n)$, after which the dominated hypervolume calculation itself is done in $O(n)$ by iterating over each point and performing an $O(1)$ calculation using the points that come before and after it in the sorting order. When calculating $S_{minus}$, the points for which the dominated hypervolume is to be calculated come from $P$, which was already sorted. This brings the complexity of this step down to $O(n)$, but it can be brought down to $O(1)$ when the order of calculations is chosen carefully, giving the algorithm a total complexity of $O(n^2)$.

The points of $P$ dominated by or equal to the lower corner of $C(a,b)$, which define $S$, are those points satisfying the following inequalities:

$$p \in P, Q^x_a.x \geq p.x, Q^y_b.y \geq p.y$$

Because of the sorting order and definition of $Q^x$ and $Q^y$, $S$ can be described equivalently as follows. The set $S$ is empty, if $a=n-b$ (the lowest value of $a$ for which $a \geq n-b$), otherwise ($a > n-b$) $S$ is formed by an uninterrupted range with $Q^x_{(n+1-b)}$ as its first element and $Q^x_a$ as its last element.
%

A row in $C_{stairs}$ is a set of cells $C_{stairs}(a,b)$ where $b$ is the same. In a single row, $S$ will always be either empty or have $Q^x_a$ as its last element. Adding 1 to $a$ adds one point to the range of points in $P$ which falls between $Q^x_{(n+1-b)}$ and  $Q^x_a$ . This makes it possible to iterate over all cells in $C_{stairs}$ while adding no more than one point to $S$ per iteration. We will do this as follows:

\begin{figure}
\begin{center}
\includegraphics[width=60mm]{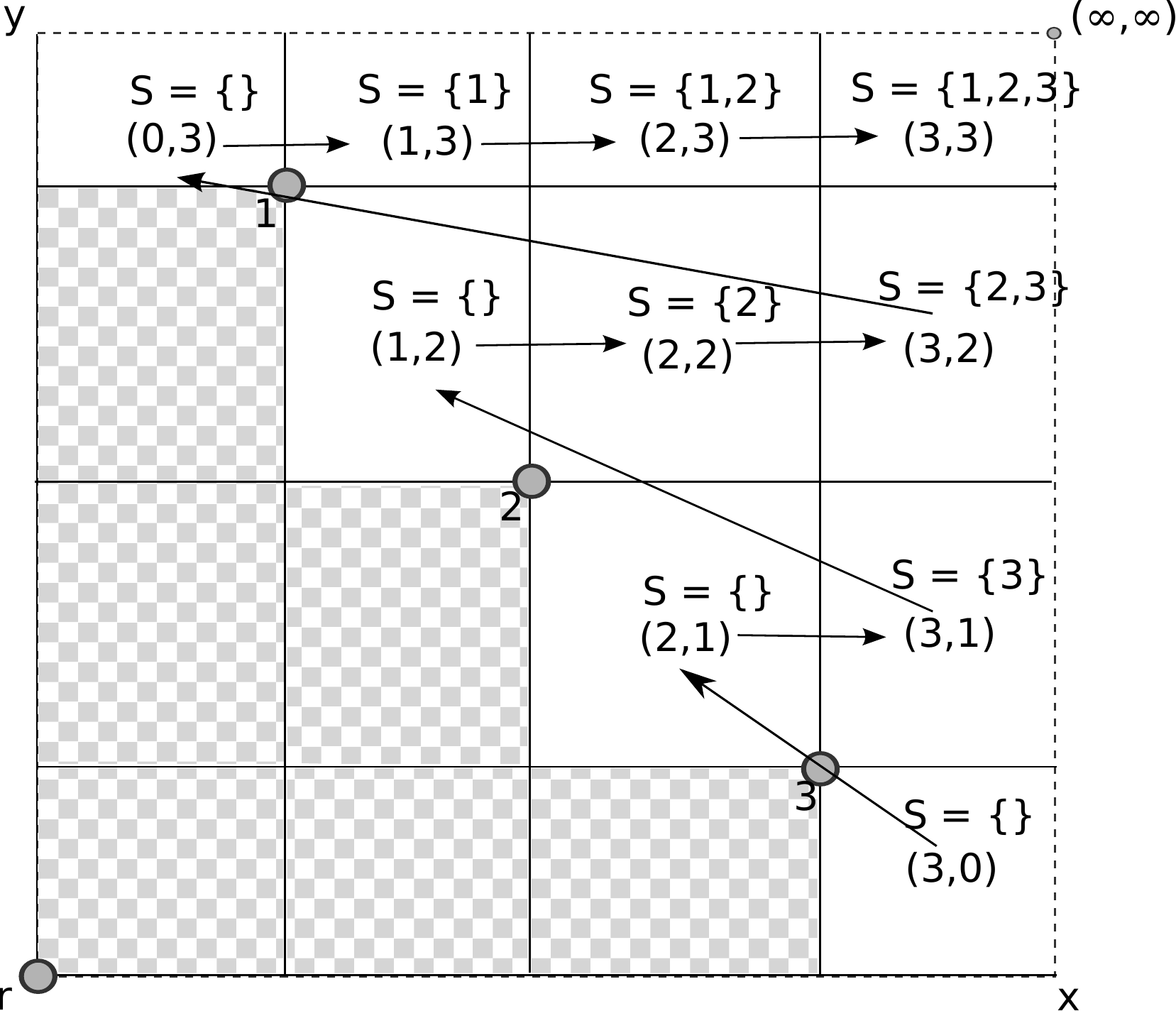}
\caption{An example showing the order of iterations which allows the hypervolume contribution of $S$ to be updated in constant time.}
\label{gridexample}
\end{center}
\end{figure}

We will start iterating over each row of $C_{stairs}$ at its first cell, where $a = n-b$. In this cell, $S = \emptyset$ and $S_{minus} = 0$.  For each iteration within a row after the first one, we add 1 to $a$ and add the point $Q^x_a$ to $S$. For an example, refer to Figure \ref{gridexample}, which shows the order of operations and the contents of $S$ during each step.

\begin{figure}
\begin{center}
\includegraphics[width=\textwidth]{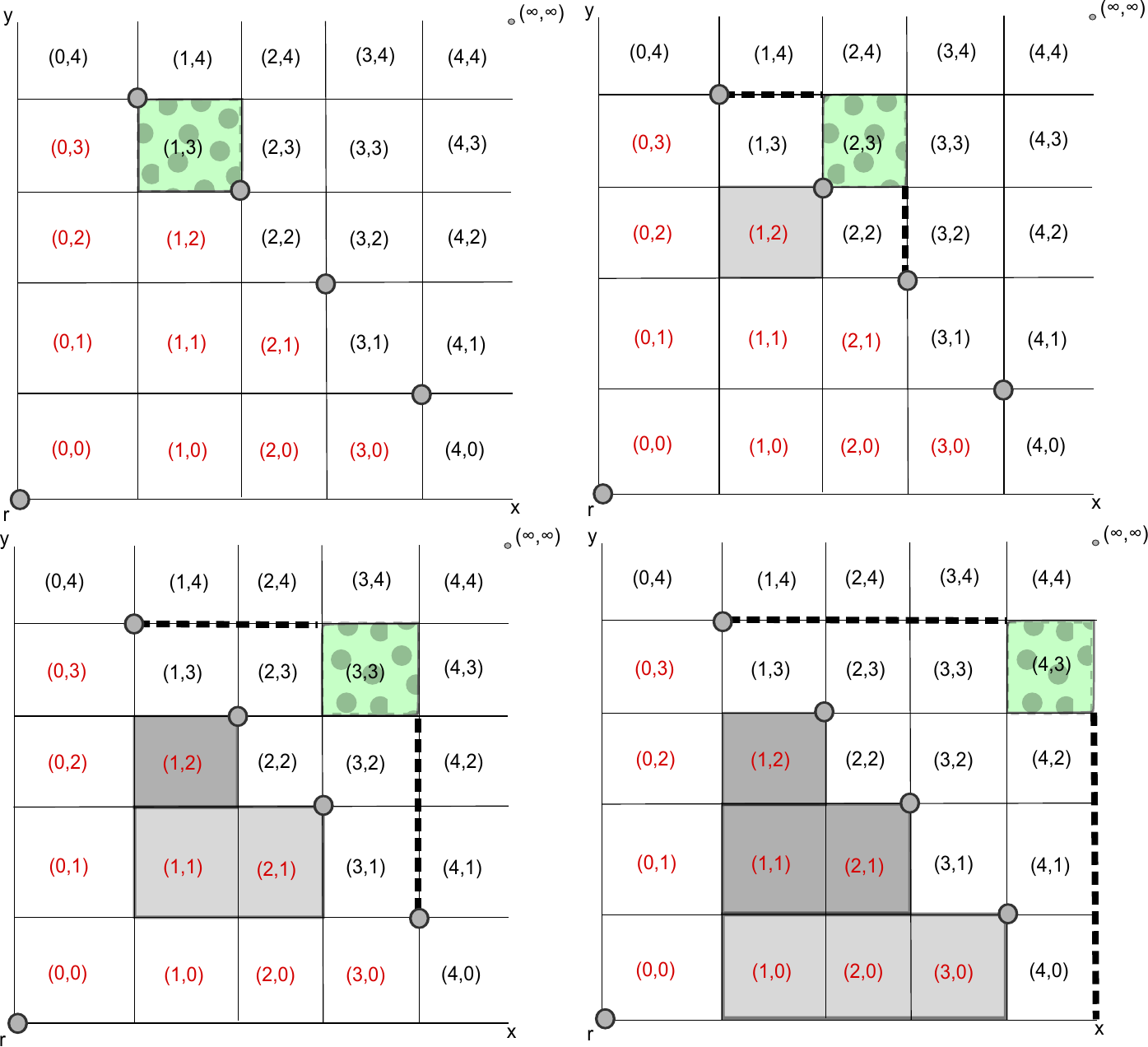}
\caption{An example showing how $S_{minus}$ changes during each iteration within a single row. The rectangular strip which is added after each iteration can be calculated with knowledge of three points: the point $Q^x_a$ is its upper corner, the point $Q^x_{a+1}$ provides the $y$ coordinate of its lower corner, and the point $Q^x_{(n-b)}$ provides the $x$ coordinate of its lower corner. Because $Q^x_{(n-b)}$ does not change, the hypervolume covered by the older points in $S$ stays the same and does not have to be re-calculated.}
\label{animationexample}
\end{center}
\end{figure}

Although the above description refers to `adding points to $S$', we only need to keep track of the first and last points of $S$ in between algorithm iterations. When a new point is added to $S$, $S_{minus}$ increases by the area covered by the rectangle from $(Q^x_{(n-b)}.x,Q^x_{a+1}.y)$ to $(Q^x_a.x, Q^x_a.y)$. Therefore, to update $S_{minus}$ after the addition of a point to $S$, only the left neighbor of the first element of $S$, the last element of $S$ and the right neighbor of the last element of $S$ are needed. Figure \ref{animationexample} shows an example of this process. This can be done in constant time in any data structure which allows the neighbors of a point to be looked up in constant time: whenever $a$ is incremented, $Q^x_a$ becomes $Q^x_{a+1}$ and $Q^x_{a+1}$ becomes $Q^x_{a+2}$. Whenever $b$ is incremented, the new $Q^x_{(n-b)}$ becomes its left neighbor, $Q^x_{(n-1-b)}$, and as we will then start iterating through values of $a$ at the beginning of the row, $Q^x_a$ becomes the new $Q^x_{(n-b)}$ as well because we have established earlier that $Q^x_a = Q^x_{(n-b)}$ in the first cell in a row of $C_{stairs}$.

We have shown that the upper bound on the complexity of determining the expected hypervolume improvement is $O(n^2)$. We can also show that the worst-case complexity can be no better than $O(n \log n)$. If the standard deviation of a candidate point's predictive distribution is 0 and the mean value vector is a point which dominates all points in $P$, then the problem of calculating its EHVI reduces to calculating the hypervolume that will be dominated by $p_{candidate}$ minus the hypervolume dominated by $P$. If it was possible to solve this calculation with lower complexity than $O(n \log n)$, then it would also be possible to reduce the calculation of $P$'s hypervolume to the problem of calculating the EHVI of a point that dominates $P$, and it has already been proven in \cite{2dhypercomplexity} that the complexity of calculating the hypervolume of a set of points in the 2-D plane is in $\Theta(n \log n)$.

\clearpage
\subsection{Empirical Performance}
As an additional verification of the correctness of the algorithm presented above, two implementations were written in C++. The first used the constant-time update scheme, and the second did not: instead of using the constant-time update scheme, $S_{minus}$ was calculated by first finding the set of points $S$ by checking each point in $P$ to see if it was dominated, and then calling a separate function on $S$ to calculate the hypervolume of this set of points.

\begin{figure}[!h]
\begin{center}
\includegraphics[width=\textwidth]{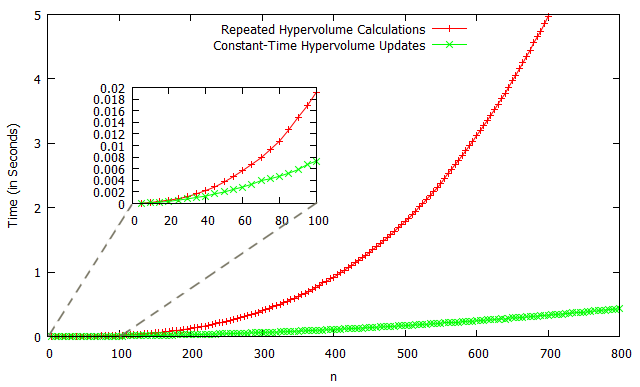}
\caption{Time needed to calculate the expected hypervolume improvement in 2-D, averaged over 10 runs. The times reported were measured on an Intel i7 quadcore CPU with 2.1 GHz clockspeed, and the code was compiled using GNU under Windows with the optimization level set to O3.}
\label{2dspeedtest}
\end{center}
\end{figure}

The expected hypervolume improvement calculated using these implementations was identical for all test problems, but their speed was not. See Figure \ref{2dspeedtest} for the empirical performance on a simple test where $P$ consisted of $n$ different points on a diagonal Pareto front. From this, it appears that using the constant-time update scheme becomes worthwhile for $n > 20$, though results might vary slightly depending on implementation and system details.

\section{Calculation of the Higher-Dimensional Expected Hypervolume Improvement}
\label{higherdimensionalehvi}
The algorithm given in \cite{exacthvi} for exactly calculating the expected hypervolume improvement is not correct when the
dimensions is higher than 2. This is because the shape of the hypervolume improvement becomes more complex when the number of dimensions increases. We will therefore derive a new formula by first decomposing the calculation into parts with less complex shapes, and then simplifying the resulting formula for the sake of more convenient calculation.

\subsection{Decomposition into Cells}
\label{3dplusformula}
In higher dimensions, the search space can be divided into cells the same way it is done in two dimensions, except instead of the boundaries being given by lines going through the points in $P$ and the reference point $r$, now the cells are separated from each other by $(m - 1)$-dimensional hyperplanes (where $m$ is the number of objective functions).

Each cell is denoted by $C(a_1,a_2,\ldots,a_m)$ where $a_1$ through $a_m$ are  integers  from $0$ to $|P|$ denoting the labeling of the cell. Then the left lower corner $\mathbf{l}$ and right upper corner $\mathbf{u}$ of the cell with label $a_1$, ..., $a_m$ are defined as follows:
Let $P' = \{r\} \cup P \cup (\infty, ..., \infty)^T$ and let $s_d[0], \dots, s_d[|P|+1]$ denote the $d$-th components of the vectors in $P'$ sorted in ascending order. Then
$l_d = s_d[a_d]$  and $u_d = s_d[a_{d}+1]$ for $d=1, \dots, m$.
In other words, corners of this cell complex are given as the intersection points of all axis-parallel $m-1$ dimensional hyperplanes through points in $P'$.



\begin{figure}
\begin{center}
\includegraphics[width=60mm]{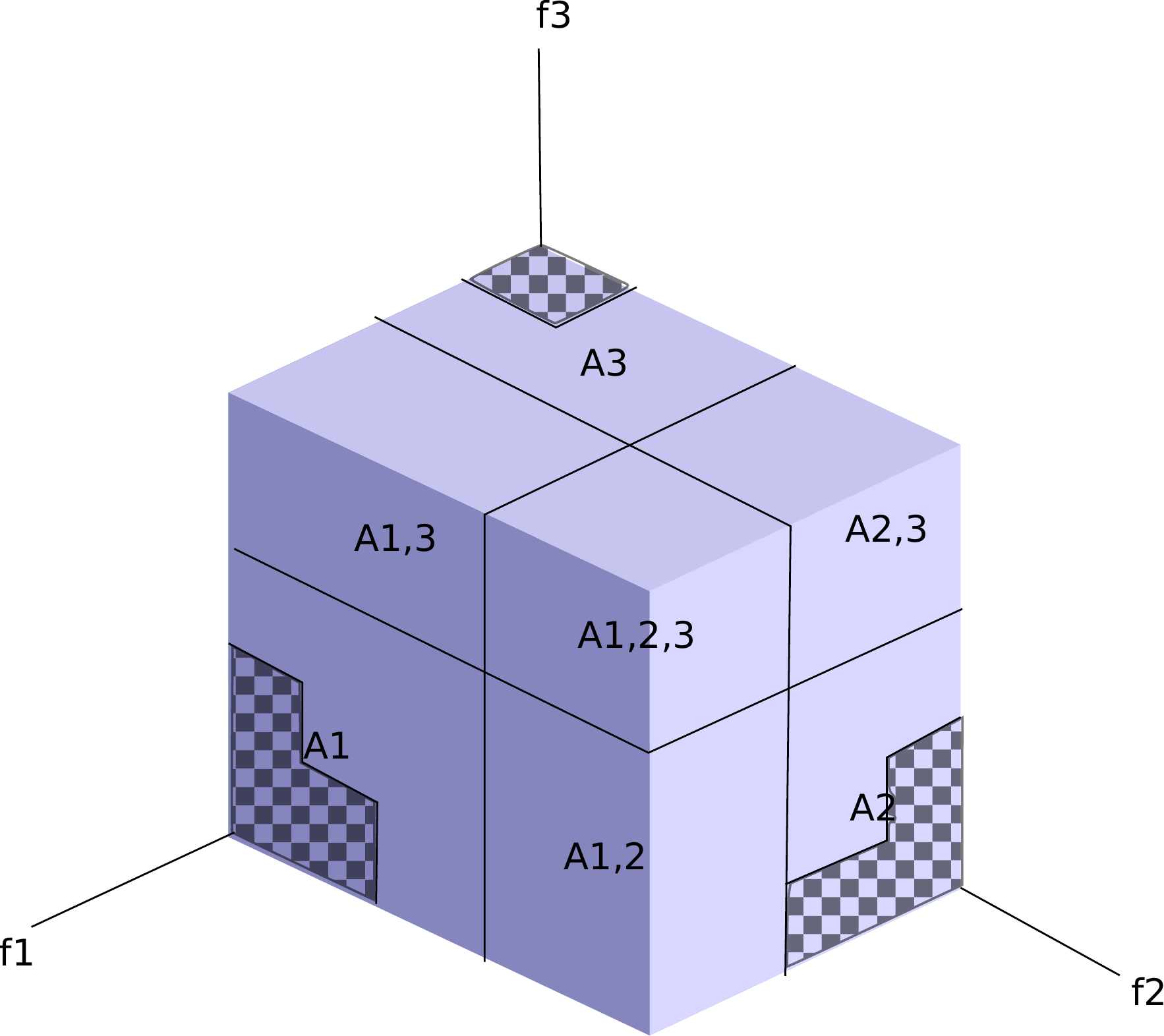}
\caption{An example showing how the quantities $A_C$ for
$C  \subseteq \{1, 2, 3\}$
are defined in a three-dimensional objective space. $A_\emptyset$ is hidden within the rectangular volume. The checkered volumes represent the volume dominated by the points in the Pareto front approximation.}
\label{ACpic}
\end{center}
\end{figure}

The hypervolume improvement of a new point $p$ with respect to the current Pareto front approximation $P$ is given by the function $\mbox{HyperVolume}\left(A \setminus \mbox{DomSet}(P)\right)$, where $A$ is the dominated hypervolume covered by $p$. This is the same as calculating $\mbox{HyperVolume}(A) - \mbox{HyperVolume}\left(\mbox{DomSet}(P) \cap A\right)$. We will denote the set of dimensions by $D = \{1,2,\ldots,m\}$. We can decompose the calculation of the hypervolume improvement of a point $p \in C(a_1,a_2,\ldots,a_m)$ as follows:

\begin{equation*}
\begin{split}
& \mbox{HI}(p) = \sum_{C \subseteq D} I_C, \ \mbox{where}\\
& I_C := \mbox{HyperVolume}\left(A_C\right) - \mbox{HyperVolume}\left(\mbox{DomSet}(P) \cap A_C\right)
\end{split}
\end{equation*}
and
$A_C$ are given by:
\begin{equation*}
\begin{split}
& A_C := \left[\begin{pmatrix}v_1 \\v_2 \\ \vdots \\ v_m \end{pmatrix}, \begin{pmatrix}w_1 \\w_2 \\ \vdots \\ w_m \end{pmatrix}\right] \\
& v_d = \begin{cases} l_d & \mbox{if }d \in C \\ r_d &\mbox{if } d \notin C \end{cases} \\
& w_d = \begin{cases} p_d &\mbox{if } d \in C \\ l_d &\mbox{if } d \notin C \end{cases}
\end{split}
\end{equation*}
See Figure \ref{ACpic} for an example in 3 dimensions.

In the above formula $\mbox{Hypervolume}$ denotes the Lebesgue measure of $\mathbb{R}^m$. Note that it can happen that the dimension of $A_C$ is strictly less than m. In this case $\mbox{Hypervolume}(A_C)=0$. We can make a similar remark about $\mbox{Hypervolume}($ $\mbox{domSet}(P) \cap A_C)$.

The values of $r_d$ and $l_d$ are constant for all points that fall within a given interval box (cell): $r$ is the reference point and is, of course, always constant, while $l$ represents the position of the lower corner of the cell. From this, it follows that $I_C$ represents the portion of the hypervolume improvement which is constant with regards to the values of $p_d, d \notin C$, and which is variable with regards to the values of $p_d, d \in C$. In fact, it is \emph{linearly related} to these values. This is a direct consequence of the way the cell boundaries are defined:

Let $\mbox{Sec}_C$ be the cross-section of $\mbox{DomSet}(P) \cap A_C$ which goes through $p$. This cross-section is defined by a projection to the dimensions not in $C$ (if $C$ consists of $k$ dimensions, the slice will be $(m-k)$-dimensional as a result). The projection of $\mbox{DomSet}(P)$ uses only those points in $P$ for which the function values in the dimensions given by $C$ are larger than the corresponding function values of $p$. We shall call this selection $P'$. No points in $P$ can fall between cell boundaries in any dimension, so the composition of $P'$ must be the same for all points within a cell. The projection of $A_C$ to the dimensions not in $C$ is constant for all points within a cell as well, because the coordinates defining $A_C$ are independent of $p$ in all dimensions not in $C$. $\mbox{HyperVolume}(\mbox{Sec}_C)$ is constant as a result -- note that here $\mbox{Hypervolume}$ is the Lebesque measure of $\mathbb{R}^{m-k}$. Because $A_C$ does not span across cell boundaries in the dimensions in $C$, $\mbox{HyperVolume}(\mbox{DomSet}(P) \cap A_C)$ is equal to the hypervolume of $\mbox{Sec}_C$ multiplied by the length of $A_C$ in each dimension in $C$, and those lengths are given by $(p_d - l_d)$ with $d \in C$.

There is one quantity $I_C$ for which $C = D$. This quantity $I_D$ is special because it is linearly related to all values of $p$. $I_D$ falls entirely within the cell, and as such, instead of projecting $P$ onto a zero-dimensional space, it can simply be said that $\mbox{HyperVolume}(\mbox{DomSet}(P) \cap A_D) = \mbox{HyperVolume}(A_D)$ if the cell is not dominated, and $\mbox{HyperVolume}(A_D \cap \mbox{DomSet}(P)) = 0$ if it is. Therefore, $I_D = \mbox{HyperVolume}(A_D)$ for non-dominated cells.

By decomposing the calculation of the hypervolume improvement, we can use the sum rule to decompose the calculation of a cell's contribution to the EHVI as well.
\begin{equation*}
\begin{split}
\int_{p = l}^{u} \sum_{C \subseteq D} I_C \cdot {PDF}(p) \mathrm{d}p = \sum_{C \subseteq D} \int_{p = l}^{u} I_C \cdot {PDF}(p) \mathrm{d}p
\end{split}
\end{equation*}

$I_C$ is calculated as the product of a constant and a set of values which are linearly related to exactly one coordinate of $p$, therefore we can first factor out the calculation of the constant part. The $PDF$ consists of independent normal distributions, allowing the probability distributions for dimensions not in $C$ (in which $I_C$ is constant) to be factured out as well. An integral consisting solely of a normal distribution can be exactly calculated using the cumulative probability distribution function $\Phi$ to calculate the probability that a point is within range of the cell.

$$\int_{p = l}^{u} I_C \cdot {PDF}(p) \mathrm{d}p =\\
I_{C}^{const} \cdot \int_{p_C = l_C}^{u_C}  \prod_{c \in C} \left( p_c - l_c \right) \cdot \prod_{c \in C} \phi_{c} \mathrm{d}p_C \cdot \prod_{c \notin C} \left( \Phi_c(u_c) - \Phi_c(l_c) \right)$$

The integral that remains is a box-shaped expected improvement where each dimension is independent. Fubini's theorem \cite{fubini} states that iterated integration, performed in any order, can be used to calculate a multiple integral under the condition that the multiple integral is absolutely convergent. The partial integrals making up the cell's contribution to the EHVI all converge to finite numbers, so we can safely use iterated integration. The result is a product of expected improvements, which are captured in the $\psi$ function described earlier:
\begin{eqnarray}
\int_{p = l}^{u} I_C \cdot {PDF}(p) \mathrm{d}p &=&\\  I_{C}^{const} \cdot \prod_{c \in C} \left( \psi(l_c, l_c, \mu_c, \sigma_c) - \psi(l_c, u_c, \mu_c, \sigma_c) \right) \cdot \prod_{c \notin C} \left( \Phi_c(u_c) - \Phi_c(l_c) \right)&&
\end{eqnarray}

It is already possible to calculate the contribution of a cell to the EHVI by summing all these terms, but the calculation can be made a bit more efficient when instead of decomposing the calculation of the hypervolume, we instead only decompose the calculation of the dominated hypervolume. In Section \ref{boringproof}, that possibility will be examined in more detail by looking at the 3-D case as an example.

\subsection{Calculation of the 3-D Expected Hypervolume Improvement}
\label{boringproof}


Consider that, in the 2-D case, we are able to calculate the hypervolume by integrating over a box bounded by the dominated hypervolume and subtracting a correction term $S_{minus}$. In higher dimensions, the correction term is not a constant, but we can still make use of a modified version of this technique. The hypervolume improvement $HI(p)$ is decomposed as follows, in three dimensions:

\begin{equation*}
\mbox{HI}(p) = I_\emptyset + I_x + I_y + I_z + I_{xy} + I_{xz} + I_{yz} + I_{xyz}
\end{equation*}

Together, they form the volume of $\left( \left[\begin{pmatrix}r_x \\r_y \\ r_z \end{pmatrix}, \begin{pmatrix}p_x \\p_y \\ p_z \end{pmatrix}\right] \setminus \mbox{DomSet}(P) \right)$. Instead of writing ${HI}(p)$ as a sum of hypervolume improvements, we can also write it as a single rectangular volume from which a dominated hypervolume is subtracted:

\begin{equation*}
\mbox{HI}(p) = \mbox{Vol}\left( \left[\begin{pmatrix}r_x \\r_y \\ r_z \end{pmatrix}, \begin{pmatrix}p_x \\p_y \\ p_z \end{pmatrix}\right]\right)
- \mbox{Vol}\left(\mbox{DomSet}(P) \cap \left[\begin{pmatrix}r_x \\r_y \\ r_z \end{pmatrix}, \begin{pmatrix}p_x \\p_y \\ p_z \end{pmatrix}\right]\right)
\end{equation*}

We can then decompose the calculation of the dominated hypervolume instead of the calculation of the hypervolume improvement. In the following decomposition of the total subtracted dominated hypervolume $S^-$, each part $S^-_C$ is equal to the subtracted dominated hypervolume needed to calculate $I_C$. When $p$ is within the integration cell bounded from below by $l$, we get the following:

\begin{equation*}
\begin{split}
S^-&= S^-_\emptyset + S^-_x + S^-_y + S^-_z + S^-_{xy} + S^-_{xz} + S^-_{yz}\\
&=
\mbox{Vol}\left(\mbox{DomSet}(P) \cap \left[\begin{pmatrix}r_x \\r_y \\ r_z \end{pmatrix}, \begin{pmatrix}l_x \\l_y \\ l_z \end{pmatrix}\right]\right)\\
&+ (p_x - l_x) \cdot \mbox{Area}\left(\mbox{DomSet}\left( \pi_{yz}\left( \sigma_{x > l_x}(P) \right) \right) \cap \left[\begin{pmatrix}r_y \\ r_z \end{pmatrix}, \begin{pmatrix}p_y \\ p_z \end{pmatrix}\right]\right)\\
&+ \ldots \\
&+ (p_x - l_x) \cdot (p_y - l_y) \cdot \left(\mbox{Max}( r_z, \pi_z ( \sigma_{x > l_x, y > l_y}(P) ) ) - r_z \right) \\
&+ \ldots \\
\end{split}
\end{equation*}
By abuse of language we use $(\mbox{Max}( r_z, \pi_z ( \sigma_{x > l_x, y > l_y}(P) ) )$ instead of the following correct notation: 
$\mbox{Max} (\{ r_z\} \cup \pi_z ( \sigma_{x > l_x, y > l_y}(P)  )$. Similar notations are also used in the sequel.   

The first thing to note is that if $r_z \geq \mbox{Max}( r_z, \pi_z ( \sigma_{x > l_x, y > l_y}(P) ) )$, $S^-_{xy} = 0$. The analogous cases are true for $S^-_{xz}$ and $S^-_{yz}$, allowing us to define a point $v$ for which, if $r = v$, all three quantities are 0:

$$v = \left[\begin{pmatrix}\mbox{Max}( r_x, \pi_x ( \sigma_{y > l_y, z > l_z}(P) ) ) \\ \mbox{Max}( r_y, \pi_y ( \sigma_{x > l_x, z > l_z}(P) ) ) \\ \mbox{Max}( r_z, \pi_z ( \sigma_{x > l_x, y > l_y}(P) ) ) \end{pmatrix} \right]$$

 The bounding box bounded by $v$ from below and $p$ from above contains the entire volume of ${HI}(p)$. This allows us to use $v$ in place of $r$ and rewrite our initial equation in a way that reduces the number of components from 8 to 5:

\begin{equation*}
\begin{split}
\mbox{HI}(p) &= \mbox{Vol}\left( \left[\begin{pmatrix}v_x \\v_y \\ v_z \end{pmatrix}, \begin{pmatrix}p_x \\p_y \\ p_z \end{pmatrix}\right]\right) \\
&- \mbox{Vol}\left(\mbox{DomSet}(P) \cap \left[\begin{pmatrix}v_x \\v_y \\ v_z \end{pmatrix}, \begin{pmatrix}l_x \\l_y \\ l_z \end{pmatrix}\right]\right)\\
&- (p_x - l_x) \cdot \mbox{Area}\left({DomSet}\left( \pi_{yz}\left( \sigma_{x > l_x}(P) \right) \right) \cap \left[\begin{pmatrix}v_y \\ v_z \end{pmatrix}, \begin{pmatrix}l_y \\ l_z \end{pmatrix}\right]\right)\\
&- (p_y - l_y) \cdot \mbox{Area}\left({DomSet}\left( \pi_{xz}\left( \sigma_{y > l_y}(P) \right) \right) \cap \left[\begin{pmatrix}v_x \\ v_z \end{pmatrix}, \begin{pmatrix}l_x \\ l_z \end{pmatrix}\right]\right)\\
&- (p_z - l_z) \cdot \mbox{Area}\left({DomSet}\left( \pi_{xy}\left( \sigma_{z > l_z}(P) \right) \right) \cap \left[\begin{pmatrix}v_x \\ v_y \end{pmatrix}, \begin{pmatrix}l_x \\ l_y \end{pmatrix}\right]\right)\\
\end{split}
\end{equation*}

The component of the EHVI integral corresponding to $\mbox{Vol}\left( \left[\begin{pmatrix}v_x \\v_y \\ v_z \end{pmatrix}, \begin{pmatrix}p_x \\p_y \\ p_z \end{pmatrix}\right]\right)$ is the only component in this equation which is variable in more than one dimension, but since it is a rectangular volume, it is simply a product of one-dimensional improvements:

$$\prod_{c \in \{x,y,z\}} \left( \psi(v_c, l_c, \mu_c, \sigma_c) - \psi(v_c, u_c, \mu_c, \sigma_c) \right)$$

$S^-_\emptyset$ is a constant. We will keep using the notation $S_{***}$, although now we will use $v$ as a reference point. Even without examining the corresponding integral it is clear that it only needs to be multiplied with the probability that a given point is within the cell. The formula for calculating this correction term is:

$$S^-_\emptyset \cdot \prod_{c \in \{x,y,z\} } (\Phi_c(u_c) - \Phi_c(l_c))$$

$S^-_x$, $S^-_y$, and $S^-_z$ are not constants, but they are each linearly related to only one coordinate of $p$. We will look at $S^-_x$ as an example:

The constant part of $S^-_x$ is $\mbox{Area}\left(\mbox{DomSet}\left( \pi_{yz}\left( \sigma_{x > l_x}(P) \right) \right) \cap \left[\begin{pmatrix}v_y \\ v_z \end{pmatrix}, \begin{pmatrix}l_y \\ l_z \end{pmatrix}\right]\right)$. This has to be multiplied by $(p_x - l_x)$. The expected value of $S^-_x$ is therefore equal to a constant multiplied by the partial expected improvement of $p_x$ over the interval $[l_x,u_x)$. This is given by:

$$\int_{p_x = l_x}^{u_x} (p_x - l_x) \phi_x(p_x)\, \mathrm{d}p_x = \psi(l_x, l_x, \mu_x, \sigma_x) - \psi(l_x, u_x, \mu_x, \sigma_x)$$

Using a new call to $\psi$ to calculate this term is not necessary. We can use the fact that $\psi$ represents the function of a one-dimensional expected improvement over a certain range bounded from below. The partial expected improvement for the region below the lower cell bound $l$ is a constant term multiplied by the chance of being within the cell's range, which is captured in the equation below:

\begin{equation*}
\begin{split}\psi(v_c, l_c, \mu_c, \sigma_c) - \psi(v_c, u_c, \mu_c, \sigma_c) =&\,\, \psi(l_c, l_c, \mu_c, \sigma_c) - \psi(l_c, u_c, \mu_c, \sigma_c)\\
& + (\Phi_c(u_c) - \Phi_c(l_c)) \cdot (l_c - v_c)
\end{split}
\end{equation*}

Both $(\Phi_c(u_c) - \Phi_c(l_c)) \cdot (l_c - v_c)$ and $\psi(v_c, l_c, \mu_c, \sigma_c) - \psi(v_c, u_c, \mu_c, \sigma_c)$ were calculated earlier, so we can reuse them to easily find $\psi(l_c, l_c, \mu_c, \sigma_c) - \psi(l_c, u_c, \mu_c, \sigma_c)$. This means that the formula for calculating the partial expected hypervolume improvement of a cell will look like the following if the cell is not dominated:

\begin{equation*}
\begin{split}
& \mbox{Let}\,\,\,\,\,\,\,\Delta\psi_c := \psi(v_c, l_c, \mu_c, \sigma_c) - \psi(v_c, u_c, \mu_c, \sigma_c), \, {c \in \{x,y,z\}} \\
& \mbox{and}\,\,\,\,\,\,\,\Delta\Phi_c := \Phi_c(u_c) - \Phi_c(l_c), \, {c \in \{x,y,z\}}
\end{split}
\end{equation*}

\begin{equation*}
\begin{split}
& - {Vol}\left({DomSet}(P) \cap \left[\begin{pmatrix}v_x \\v_y \\ v_z \end{pmatrix}, \begin{pmatrix}l_x \\l_y \\ l_z \end{pmatrix}\right]\right) \cdot \prod_{c \in \{x,y,z\}} \Delta\Phi_c\\
& - (\Delta\psi_z - \Delta\Phi_z \cdot (l_z - v_z)) \\
& \,\,\,\,\, \cdot {Area}\left({DomSet}\left( \pi_{yz}\left( \sigma_{x > l_x}(P) \right) \right) \cap \left[\begin{pmatrix}v_y \\ v_z \end{pmatrix}, \begin{pmatrix}l_y \\ l_z \end{pmatrix}\right]\right) \cdot \prod_{c \in \{x,y\}} \Delta\Phi_c\\
&- (\Delta\psi_y - \Delta\Phi_y \cdot (l_y - v_y)) \\
& \,\,\,\,\, \cdot {Area}\left({DomSet}\left( \pi_{xz}\left( \sigma_{y > l_y}(P) \right) \right) \cap \left[\begin{pmatrix}v_x \\ v_z \end{pmatrix}, \begin{pmatrix}l_x \\ l_z \end{pmatrix}\right]\right)  \cdot \prod_{c \in \{x,z\}} \Delta\Phi_c \\
& - (\Delta\psi_x - \Delta\Phi_x \cdot (l_x - v_x))  \\
& \,\,\,\,\, \cdot{Area}\left({DomSet}\left( \pi_{xy}\left( \sigma_{z > l_z}(P) \right) \right) \cap \left[\begin{pmatrix}v_x \\ v_y \end{pmatrix}, \begin{pmatrix}l_x \\ l_y \end{pmatrix}\right]\right) \cdot \prod_{c \in \{y,z\}} \Delta\Phi_c
\end{split}
\end{equation*}

And it will be 0 otherwise.

\subsection{Simple Higher Dimensional Expected Hypervolume Improvement Computation}
\label{awesomeproof}
Although we are currently decomposing our integral into different quantities in order to calculate it, we can also calculate the sum of these quantities using a single dominated hypervolume calculation, though this has downsides which will be explored later. This subsection will give the general formula for doing so. Recall how we decomposed the calculation of the hypervolume improvement in Section \ref{3dplusformula}:

\begin{equation*}
\begin{split}
& \mbox{HI}(p) = \sum_{C \subseteq D} I_C\\
& I_C = \mbox{HyperVolume}(A_C) - \mbox{HyperVolume}(\mbox{DomSet}(P) \cap A_C)
\end{split}
\end{equation*}

This sum can be rearranged to the following:

$$\sum_{C \subseteq D}  \mbox{HyperVolume}(A_C) - \sum_{C \subseteq D} \mbox{HyperVolume}(\mbox{DomSet}(P) \cap A_C)$$

Since the quantities $A_C$ sum to a generalized rectangular volume, we could just as readily calculate the total volume of $A$ directly. This is what we did in Section \ref{boringproof}, where the correction terms $\mbox{HyperVolume}(\mbox{DomSet}(P) \cap A_C)$ were still calculated separately.
Clearly $\bigcup_{C \subseteq D} A_C = A$. Moreover the $A_Cs$ are mutually disjoint except for common
boundary points. From this we get that
$\mbox{DomSet}(P) \cap (\bigcup_{C \subseteq D} A_C) = \mbox{DomSet}(P) \cap A$
(and the $\mbox{DomSet}(P)\cap A_C$s
are mutually disjoint except for the boundaries). Thus $$\mbox{Hypervolume}(\mbox{DomSet}(P)\cap A) =\sum_{C \subseteq D} \mbox{Hypervolume}(\mbox{DomSet}(P) \cap A_C). $$
We initially decomposed the $\mbox{HI}(p)$ in this way in order to compute the corresponding $EI$-integral.


We have determined that each partial quantity $\mbox{HyperVolume}(\mbox{DomSet}(P) \cap A_C)$ depends linearly on the dimensions which its corresponding volume $A_C$ depends on, and is constant in the same dimensions in which $A_C$ is constant. This is true as well when $\mbox{HyperVolume}(\mbox{DomSet}(P) \cap A)$ is first calculated, and then split into the various volumes representing $\mbox{DomSet}(P) \cap A_C$. Because of this, we can calculate an $m$-dimensional EHVI using only a single $m$-dimensional hypervolume calculation per cell. We need to calculate the hypervolume improvement of each cell's \emph{center of mass}, $\bar p$.

$$\bar p_d = \frac{\int_{p_d = l_d}^{u_d} p_d \cdot \phi_d(p_d) \mathrm{d}p}{ \Phi_d(u_d) - \Phi_d(l_d) }$$

The integral can be calculated as if it is an expected improvement where the currently best solution is 0. However, we already need to compute $\psi(r_d, l_d, \mu_d, \sigma_d) - \psi(l_d, u_d, \mu_d, \sigma_d)$ to calculate the component of the EHVI corresponding to $A$, and the following equation holds:

 $$\frac{\psi(0, l_d, \mu_d, \sigma_d) - \psi(0, u_d, \mu_d, \sigma_d)}{\Phi_d(u_d) - \Phi_d(l_d)} = \frac{\psi(r_d, l_d, \mu_d, \sigma_d) - \psi(r_d, u_d, \mu_d, \sigma_d)}{\Phi_d(u_d) - \Phi_d(l_d)} + r_d$$

Dividing a partial expected improvement over a range $[l_d,u_d)$ by the chance of being in that range (given by $\Phi_d(u_d) - \Phi_d(l_d)$) gives the expected improvement of points which are known to lie within that range. Adding the value of $r_d$ gives the expected $d$th coordinate of a point in the objective space bounded from below by $r$.

This means that the general formula for calculating the partial expected improvement in a cell is the following if the cell is not dominated:

$${EI} = \prod_{d \in D} \left( \psi(r_d, l_d, \mu_d, \sigma_d) - \psi(r_d, u_d, \mu_d, \sigma_d) \right) - S^- \cdot \prod_{d \in D} (\Phi_d(u_d) - \Phi_d(l_d)), \mbox{ where}$$

$$S^- = \mbox{HyperVolume}\left( \mbox{DomSet}(P) \cap \left[\begin{pmatrix}r_1 \\r_2 \\ \vdots \\ r_m \end{pmatrix}, \begin{pmatrix} \bar p_1 \\ \bar p_2 \\ \vdots \\ \bar p_m \end{pmatrix}\right]\right) \mbox{ and}$$

$$\bar p_d = r_d + \frac{ \psi(r_d, l_d, \mu_d, \sigma_d) - \psi(r_d, u_d, \mu_d, \sigma_d)}{ \Phi_d(u_d) - \Phi_d(l_d)} $$

And 0 otherwise.

\subsection{Complexity and Algorithm Details}
Any algorithm which iterates over all grid cells described in Section \ref{higherdimensionalehvi} will have a time complexity of $\Omega(n^m)$. This is further increased by the complexity of the calculations within each grid cell. The algorithm of Section \ref{awesomeproof} requires an $m$-dimensional hypervolume to be calculated for each cell that is not dominated. Calculating a 3-dimensional hypervolume can be done in $O(n \log n)$, which results in a time complexity of $O(n^4 \log n)$. However, as will be shown in Section \ref{sliceupdates}, constant-time calculations within each grid cell are possible with $O(n^3)$ total preparation time, resulting in an algorithm of complexity $O(n^3)$. Similar $O(n^m)$ algorithms are conjectured to exist for $m > 3$.

One important thing to note, is that the expected hypervolume improvements for multiple individuals can be calculated at the same time without having to perform the hypervolume calculations more than once when using the decomposition described in Section \ref{boringproof}, because the hypervolume calculations are not dependent on the mean and standard deviation of the probability distribution. The algorithm described in Section \ref{awesomeproof} does not have this advantage.

\section{$O(n^3)$-time 3-D Expected Hypervolume Improvement Calculations}
\label{sliceupdates}
In Section \ref{2dcomplexity} we showed that calculating the 2-D expected hypervolume improvement is possible with time complexity $O(n^2)$. Although the algorithm described in that subsection made use of characteristics of a 2-D Pareto approximation set which are not present in higher dimensions, this subsection will show that there is also a way to calculate the 3-D EHVI with time complexity $O(n^3)$. In other words: the calculations necessary for computing the partial expected hypervolume improvement of each grid cell will be performed in constant time. The trade-off is that we will need $O(n^2)$ extra memory.

The only calculations which have a complexity higher than constant time are the dominated hypervolume calculations. If we use the simple algorithm described in Section \ref{awesomeproof}, we only need to perform a single 3-dimensional hypervolume calculation to find the correction term that we need. However, we will start out with the algorithm described in Section \ref{boringproof} (\emph{without} replacing $r$ by $v$), because it lends itself better to the re-use of old hypervolume calculations. Three sets of correction terms are needed to calculate the partial expected hypervolume improvement of a cell:
\begin{itemize}
\item $S^-_\emptyset$, a constant correction term which requires a three-dimensional hypervolume calculation.
\item $ S^-_x$, $S^-_y$ and $S^-_z$, which each require a two-dimensional hypervolume calculation. We will call the 2-D areas used in the calculation of these correction terms ${xslice}$, ${yslice}$ and ${zslice}$, respectively.
\item $S^-_{xy}$, $S^-_{xz}$ and $S^-_{yz}$, which requires a `one-dimensional' hypervolume calculation.
\end{itemize}

\begin{figure}[h]
\begin{center}
\begin{minipage}[b]{0.49\linewidth}
\includegraphics[width=\textwidth]{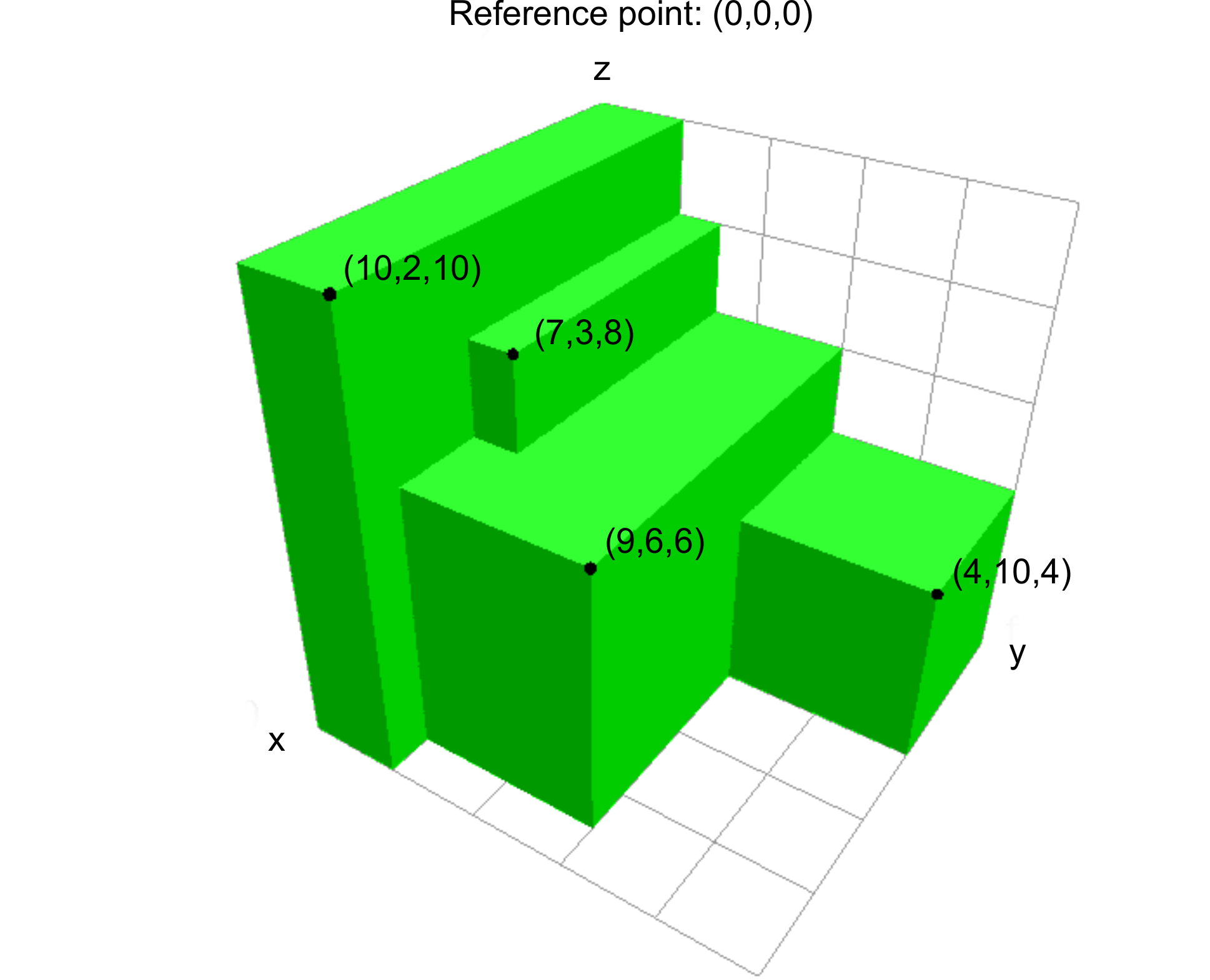}
\end{minipage}
\begin{minipage}[b]{0.39\linewidth}
\includegraphics[width=\textwidth]{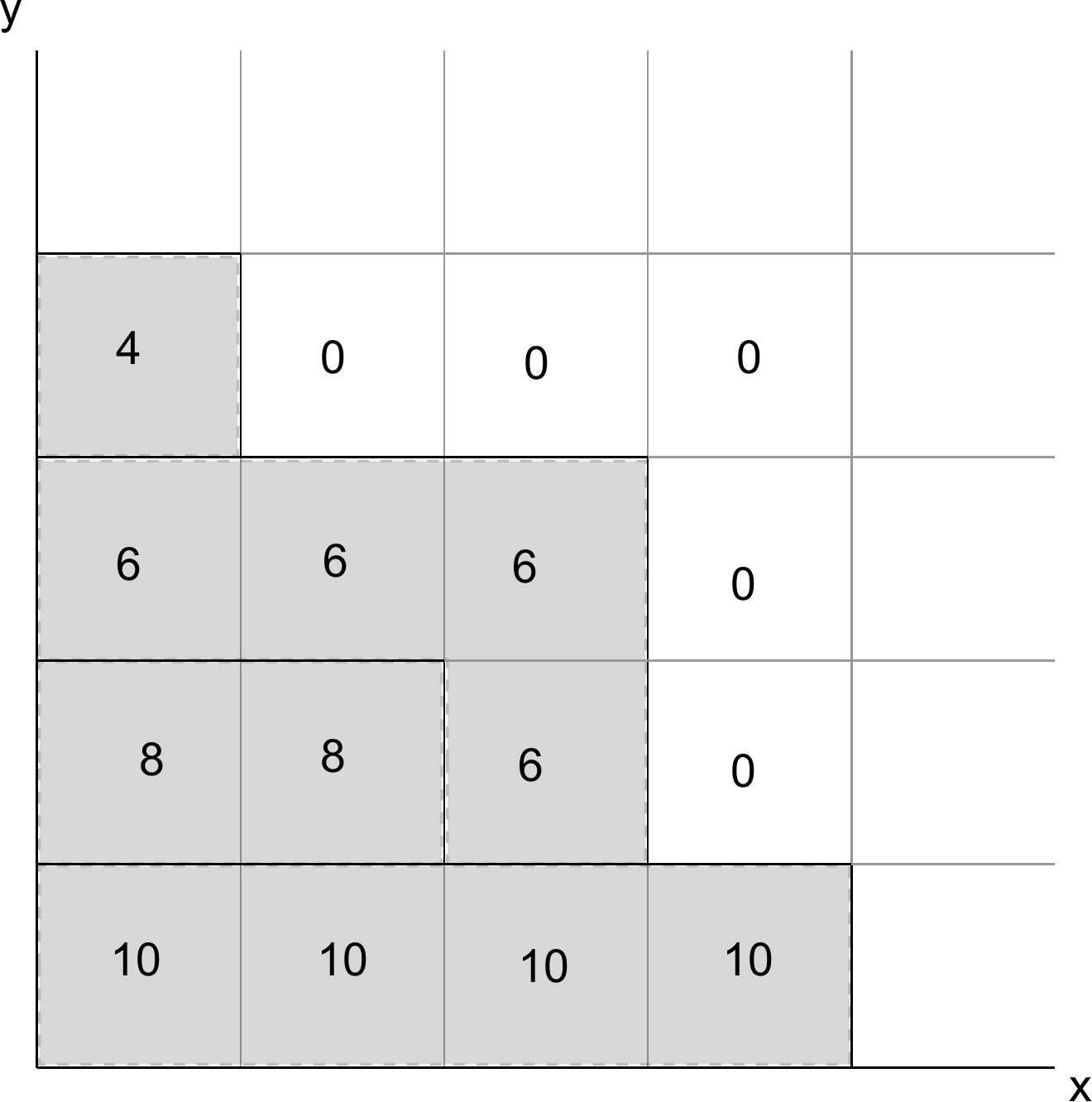}
\end{minipage}
\caption{Example height array $H_z$ for a population consisting of 4 points, which is visualized on the left. Cells on the outermost edge of the integration area (which stretch out to $\infty$ in some dimension) are always non-dominated. }
\label{Hzexample}
\end{center}
\end{figure}

Instead of calculating these correction terms afresh for each cell, it is possible to perform all necessary hypervolume calculations in only $O(n^3)$ time total. The first step is to create a data structure which allows us to see whether or not a cell is dominated in $O(1)$ time. This can simply be a two-dimensional array holding the highest value of $z$ for which the cell is dominated, which we shall call $H_z$. A simple way to fill this array is to iterate over all points $q \in P$ in order of ascending $z$ value, setting the array value $H_z(a_1,a_2)$ to $z$ if $q$ dominates the lower corner of $C(a_1,a_2,0)$.  The complexity of this operation is in $O(n^2 n + n \log n) = O(n^3)$. This only needs to be done once, so the $O(n^3)$ time complexity does not increase the total asymptotic time complexity of computing the EHVI in 3-D. Figure \ref{Hzexample} shows an example.

Besides containing information that allows constant-time evaluation of whether a cell is dominated, the value of $S^-_{xy}$ for a cell $C(a_1,a_2,a_3)$ that is not dominated is also given by $H_z(a_1,a_2)$. If we build two more height arrays $H_x$ and $H_y$ where we use the highest value of $x$ and $y$ instead of $z$, we can determine the results of all three of the one-dimensional hypervolume calculations in constant time during cell calculations.

Now, only the two-dimensional hypervolume calculations represented by ${xslice}$, ${yslice}$ and ${zslice}$, and the three-dimensional hypervolume calculation represented by $S^-_\emptyset$, still have a complexity greater than constant time. For notational simplicity, we have omitted their dependence on a particular cell from the notation until now, but in order to show the relations between correction terms of different cells, we will write `$S^-_\emptyset$ belonging to $C(a_1,a_2,a_3)$' as $C(a_1,a_2,a_3).S^-_\emptyset$, and likewise for the two-dimensional hypervolumes.

The value of $S^-_\emptyset$ is related to the values of ${xslice}$, ${yslice}$ and ${zslice}$ in the following way:
\begin{itemize}
\item $C(a_1,a_2,a_3).xslice = \frac{C(a_1+1,a_2,a_3).S^-_\emptyset - C(a_1,a_2,a_3).S^-_\emptyset}{u_x - l_x}$
\item $C(a_1,a_2,a_3).yslice =\frac{C(a_1,a_2+1,a_3).S^-_\emptyset - C(a_1,a_2,a_3).S^-_\emptyset}{u_y - l_y}$
\item $C(a_1,a_2,a_3).zslice = \frac{C(a_1,a_2,a_3+1).S^-_\emptyset - C(a_1,a_2,a_3).S^-_\emptyset}{u_z - l_z}$
\end{itemize}


With our height array $H_z$, we can calculate all values of ${zslice}$ for a given value of $a_3$ in $O(n^2)$ time. We can also calculate all values of $S^-_\emptyset$ for a given value of $a_3$ in $O(n^2)$ time, provided $a_3 = 0$ or we have both $S^-_\emptyset$ and ${zslice}$ for the cells where $a_3$ is one lower. The details of these calculations will be given below. If we go through our cells in the right order (with $a_3$ starting at 0, incrementing it only after we have performed the calculations for all cells with a given value of $a_3$), we only need to update the values of ${zslice}$ and $S^-_\emptyset$ $n$ times, resulting in an algorithm for the full computation with complexity in $O(n^3)$. If we know the value of $S^-_\emptyset$ for all cells with a given value of $a_3$, we can use the formulas given above to calculate ${xslice}$ and ${yslice}$ in constant time whenever we need them, so we do not need to calculate these constants in advance. 

The details of calculating ${zslice}$ using the height array are as follows. We will iterate through the possible values of $a_1$ and $a_2$ in ascending order. We know that ${zslice}(a_1,a_2)$ is 0 if $a_1 = 0$ or $a_2 = 0$. If our height array shows that $C(a_1-1,a_2-1,a_3)$ is dominated, ${zslice}(a_1,a_2)$ is set equal to the area of the 2-D rectangle from its lower corner to $(r_x,r_y)$. Else, if that cell is not dominated, ${zslice}(a_1,a_2)$ is set equal to ${zslice}(a_1-1,a_2) + {zslice}(a_1,a_2-1) - {zslice}(a_1-1,a_2-1)$. The value of ${zslice}(a_1-1,a_2-1)$ is removed as this is the area which is overlapping, causing it to be added twice otherwise.

\begin{figure}[Ht]
\begin{minipage}[b]{\textwidth}
\begin{center}
\begin{minipage}[b]{0.39\linewidth}
\includegraphics[width=\textwidth]{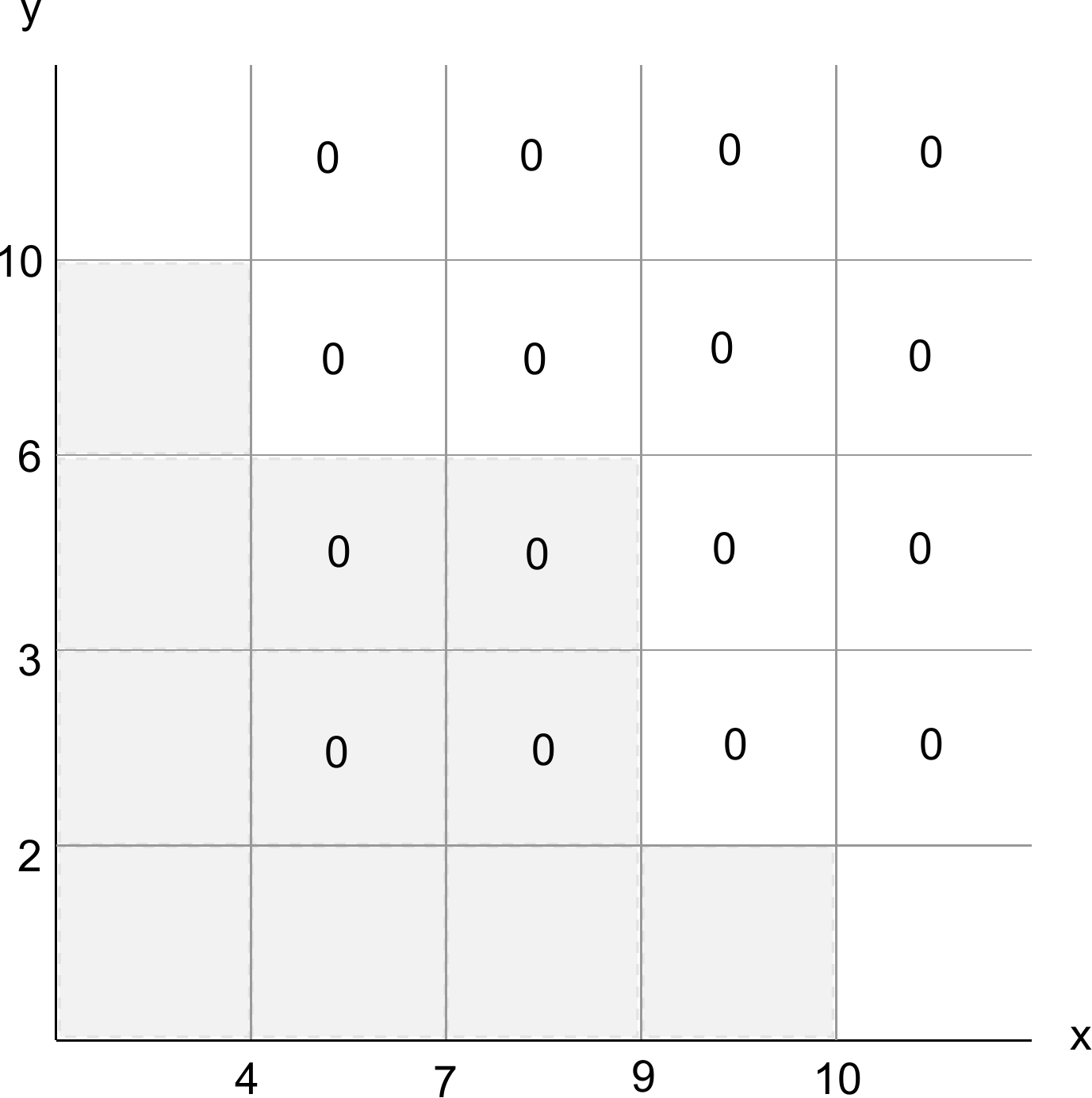}
\end{minipage}
\begin{minipage}[b]{0.39\linewidth}
\includegraphics[width=\textwidth]{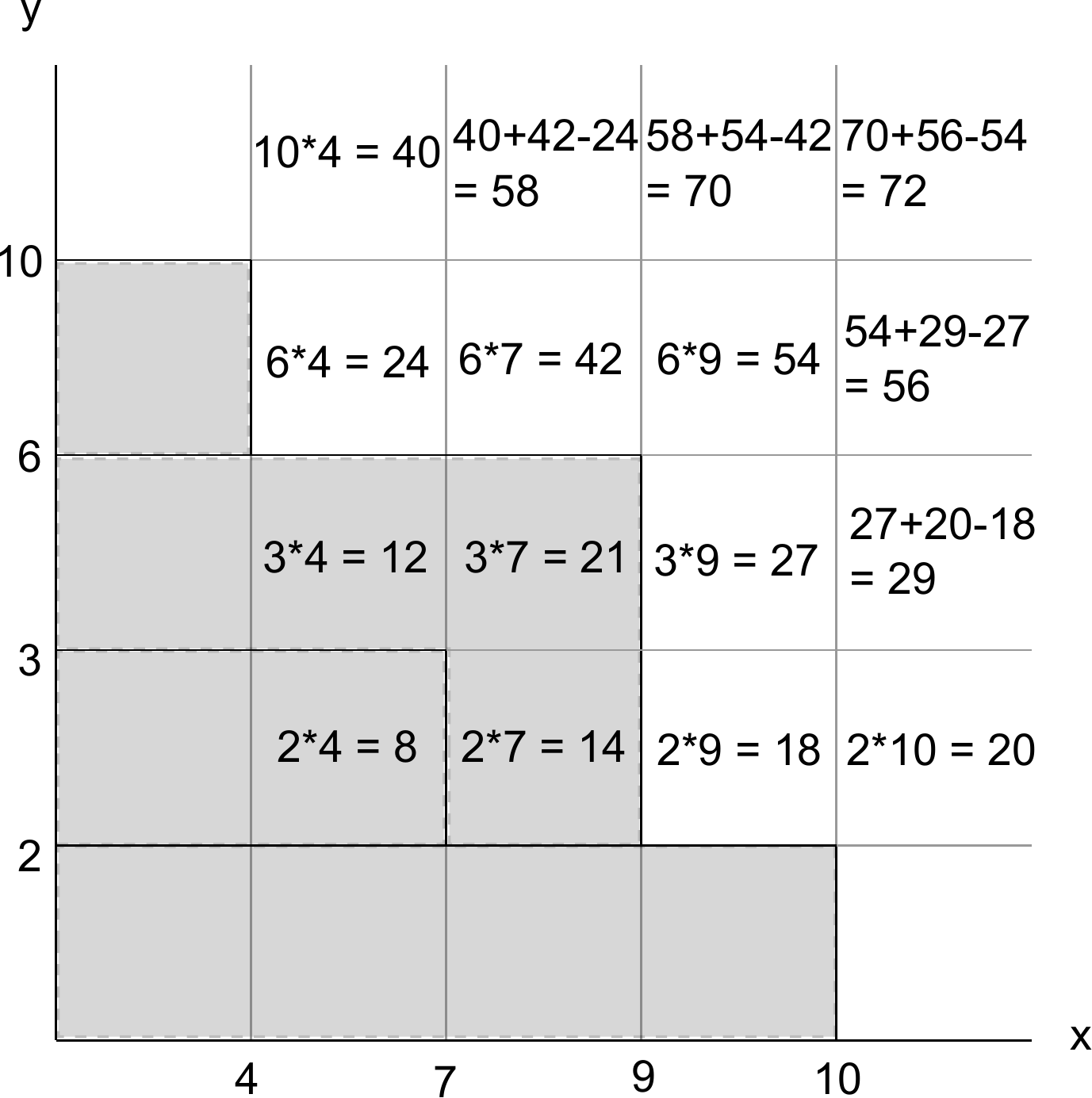}
\end{minipage}
\end{center}
\end{minipage}
\begin{minipage}[b]{\textwidth}
\begin{center}
\begin{minipage}[b]{0.39\linewidth}
\includegraphics[width=\textwidth]{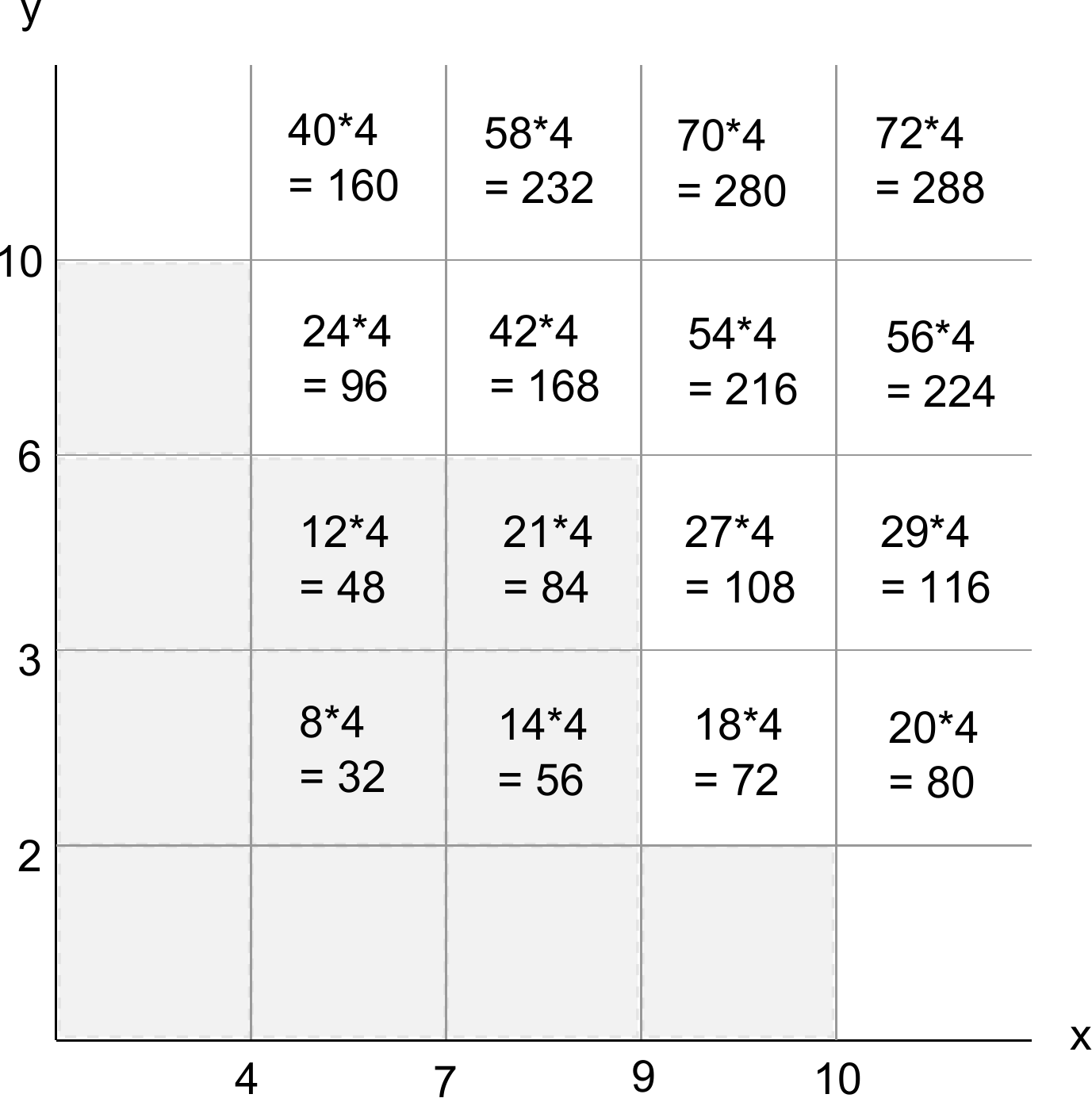}
\end{minipage}
\begin{minipage}[b]{0.39\linewidth}
\includegraphics[width=\textwidth]{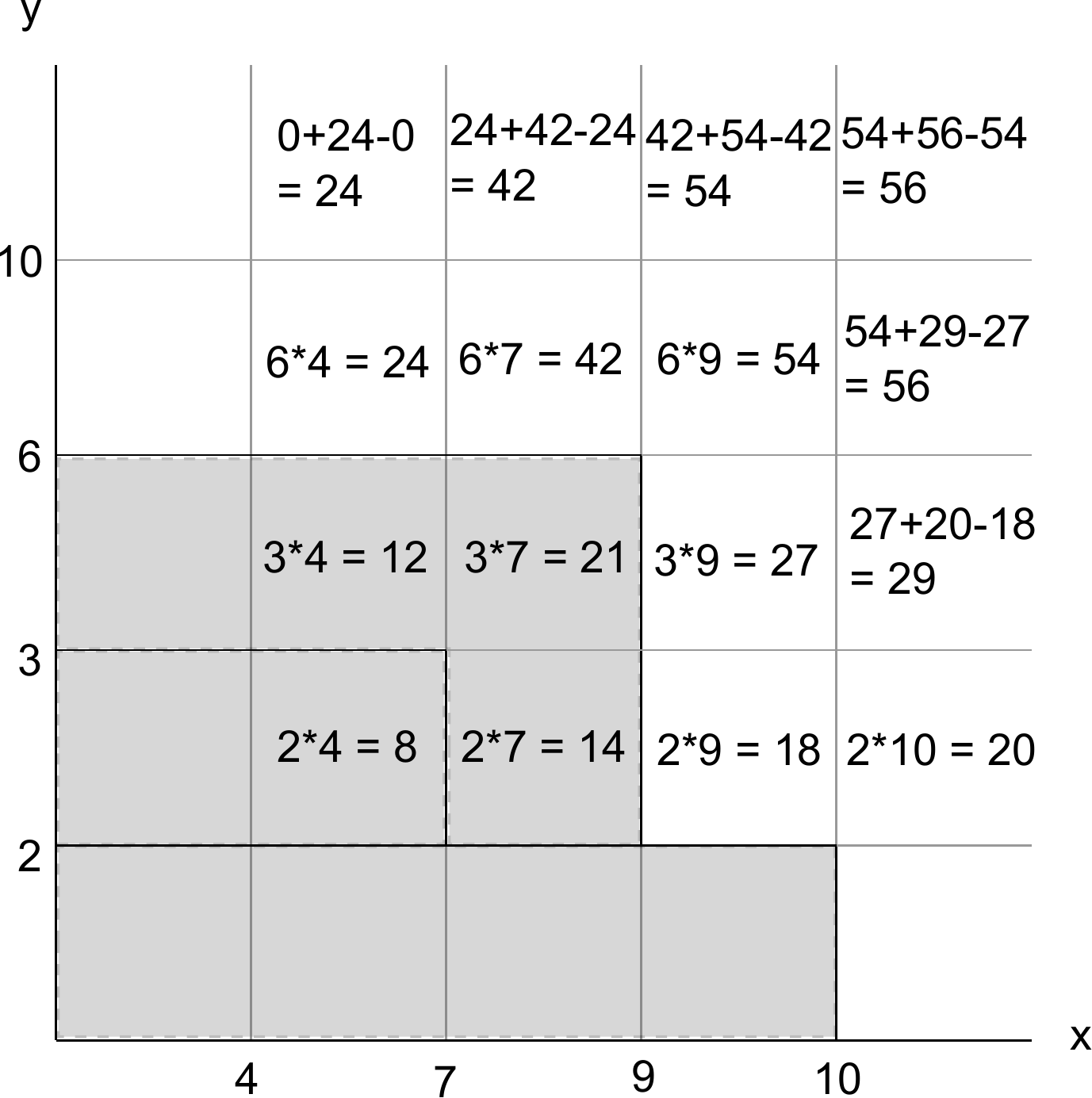}
\end{minipage}
\end{center}
\end{minipage}
\begin{minipage}[b]{\textwidth}
\begin{center}
\begin{minipage}[b]{0.39\linewidth}
\includegraphics[width=\textwidth]{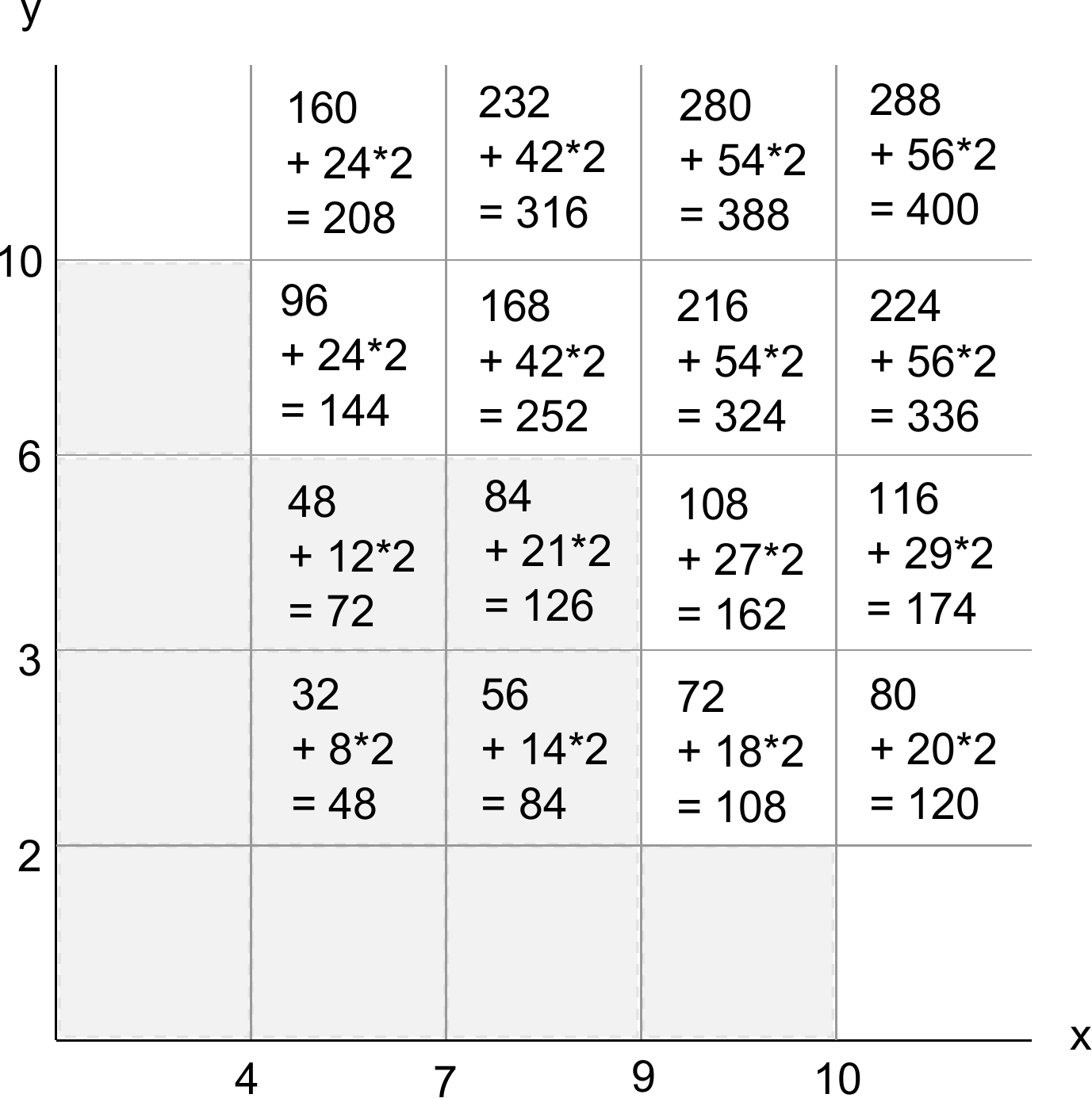}
\end{minipage}
\begin{minipage}[b]{0.39\linewidth}
\includegraphics[width=\textwidth]{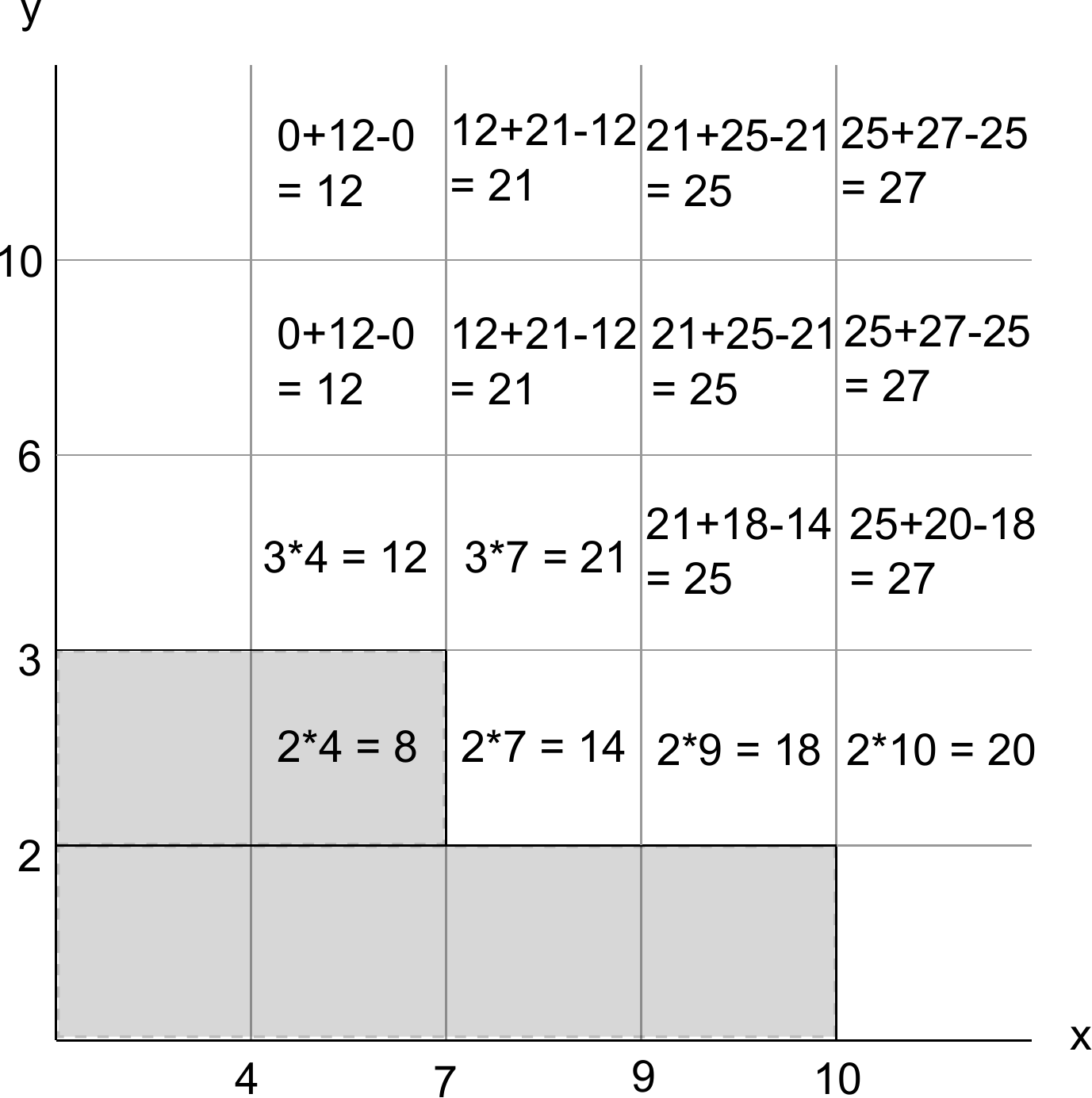}
\end{minipage}
\end{center}
\end{minipage}

\caption{Some values of ${zslice}$ and $S^-$ for the example shown in Figure \ref{Hzexample}, with $a_3 = 0$, 1 and 2, respectively. The $x$ and $y$ values of each cell's lower corner are shown on the axes. The grids with the values of $S^-$ are on the left and the grids with the values of ${zslice}$ are on the right. }
\label{zslicestuff}
\end{figure}

For an example, refer to Figure \ref{zslicestuff}.

\section{Empirical Tests and Results}
\label{emptests}
Five different implementations of a 3-D expected hypervolume improvement calculation algorithm were used throughout the following tests, referred to as the 8-term, 5-term, 2-term, slice-update and Monte Carlo schemes. The goal of comparing the exact calculation algorithms to a Monte Carlo scheme is twofold. First, by computing the Expected Hypervolume Improvement in different ways, the algorithms and their implementations will be thoroughly validated. Second, the time consumption of the algorithms will be compared. This is of particular interest because Monte Carlo schemes are often used as fast approximations to exact computations.

\begin{itemize}
\item The 8-term scheme is a direct implementation of the calculations described in Section \ref{boringproof}.

\item The 5-term scheme implements the slightly simplified calculations described in Section \ref{boringproof}.

\item The 2-term scheme implements the calculations described in Section \ref{awesomeproof}.

\item The slice-update scheme implements the algorithm described in Section \ref{sliceupdates}.

\item The Monte Carlo scheme uses Monte Carlo integration to give an approximation of the expected hypervolume improvement. Its random number generator uses the Box-Muller transform \cite{boxmuller} in combination with the Mersenne Twister algorithm \cite{twister} (specifically, the 32-bit MT19937 variant from the C++ standard library, implemented in GCC) to generate normally distributed pseudo-random numbers. Due to the nature of Monte Carlo algorithms, it is impossible to get an exact answer out of this scheme. The expected error of Monte Carlo integration is related to the number of trials $m$ by $\frac{1}{\sqrt m}$, which means that to make the estimate ten times more accurate, a hundred times more trials are required.
\end{itemize}

The implementations of $\psi$ and the Gaussian cumulative distribution function were identical for all schemes, except for the Monte Carlo scheme where they were not used. The 2-D and 3-D hypervolume calculation functions were also identical between those schemes which used them. Standard C++ library functions were used for sorting and for the implementation of the Gaussian error function \emph{erf}.

\subsection{Monte Carlo Verification}
As a verification of the correctness of the algorithms, the expected hypervolume improvements calculated by all schemes on several test problems were compared to each other and to the value which the Monte Carlo scheme converged towards.

\begin{figure}[h]
\begin{center}
\includegraphics[width=\textwidth]{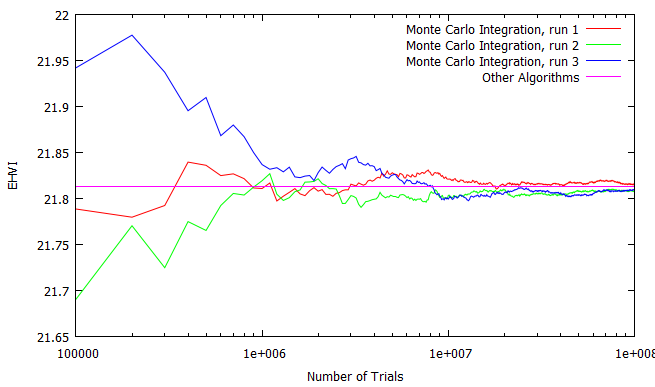}
\caption{Logarithmic-scale graph of the convergence of Monte Carlo integration. The answer was measured every 100.000 iterations.}
\label{simplemontecarlos}
\end{center}
\end{figure}

The graph in Figure \ref{simplemontecarlos} shows the results of running the algorithms on a simple test problem. The population consisted of three points: $(1,2,3)$, $(2,3,1)$ and $(3,1,2)$. The reference point was set to $(0,0,0)$. The median vector for the Gaussian distribution was set to $(3,3,3)$, placing it right between cell borders, and the standard deviation was set to $(2,2,2)$. All non-Monte Carlo schemes gave exactly identical answers, which was likely due to the simplicity of the test case, because rounding errors in the floating-point calculations would have resulted in small differences otherwise. The Monte Carlo scheme was allowed to run for 100.000.000 iterations.

\begin{figure}[h!]
\begin{center}
\begin{minipage}[b]{0.8\linewidth}
\includegraphics[width=\textwidth]{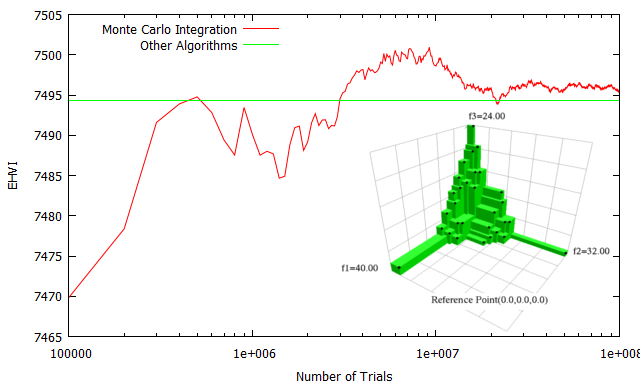}
\end{minipage}
\begin{minipage}[b]{0.8\linewidth}
\includegraphics[width=\textwidth]{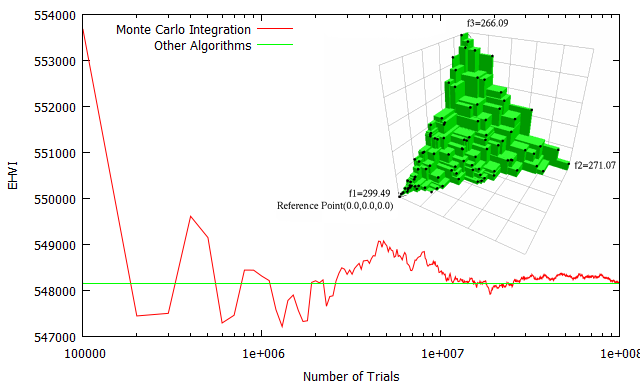}
\end{minipage}
\caption{Two logarithmic-scale graphs showing the convergence of Monte Carlo integration, along with visualizations of the Pareto approximation sets.}
\label{morecarlos}
\end{center}
\end{figure}

Figure \ref{morecarlos} shows the results of running the algorithm on a few more complex populations. The first consists of 30 points, some of which had identical values to another point in the population in one of their dimensions (creating cells of size 0). The second consists of 100 points with a bias towards one area of the search space. The results of all non-Monte Carlo schemes on these two test problems were identical to 15 and 14 digits, respectively. The double-precision floating numbers which were used in the implementations are accurate to approximately the 15th decimal, so the answers can safely be considered identical.

The convergence of the Monte Carlo integration, as well as the near-identical answers generated by the different approaches towards calculating the expected hypervolume improvement, both support the validity of the calculations described in this thesis.

\clearpage
\subsection{Empirical Performance}

\begin{figure}[h!]
\includegraphics[width=\textwidth]{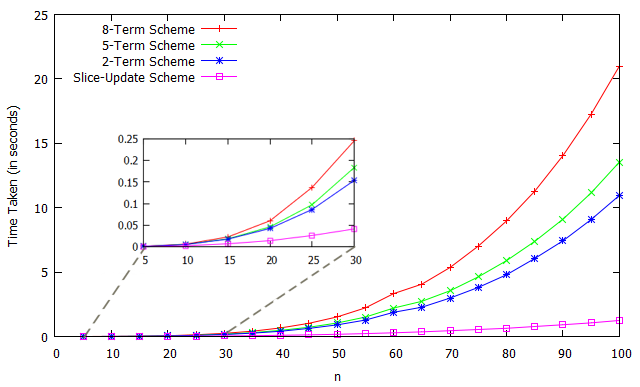}
\caption{Time needed to calculate the expected hypervolume improvement for a Pareto approximation set consisting of $n$ points randomly selected on the surface of a sphere, averaged over 10 runs.}
\label{3dspheretime}
\end{figure}

To test the empirical performance of the exact calculation schemes, they were tested on mutually non-dominated populations of varying sizes that were generated by selecting $n$ pseudo-random points which were uniformly distributed on a spherical surface. The time needed for calculating the expected hypervolume improvement was measured (along with all operations required to do so, such as sorting the populations, but not including the time needed to generate the populations). The seed of the pseudorandom generator was the same for each calculation scheme that was tested. Figure \ref{3dspheretime} shows the results. There is a noticeable difference in speed between the 8-term, 5-term and 2-term scheme, but they are in the same complexity class and for any given $n$, their performance relative to each other is roughly the same. The slice-update scheme, by contrast, performs better relative to the other schemes when $n$ increases, as would be expected due to its lower complexity. Even for small $n$ it outperforms the other algorithms.

What is interesting is that going from 8 terms to 5 terms causes a greater improvement than going from 5 terms to 2 terms, even though 2-dimensional hypervolume calculations are completely removed from the equation when going to 2 terms. No solid conclusions can be drawn from the magnitude of the differences, as they might depend on the implementation details of the code and the compiler optimizations. However, this does show that simplifying calculations can make a big difference for the speed of an algorithm. A benefit of the 2-term scheme which is not captured in the graph, is that it is the simplest scheme in terms of the number of operations that must be implemented, so the time needed to implement it will be shorter.

\begin{figure}[h!]
\begin{center}
\begin{minipage}[b]{0.49\linewidth}
\includegraphics[width=\textwidth]{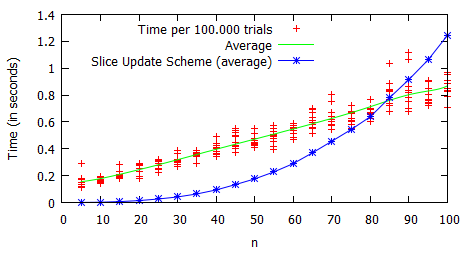}
\end{minipage}
\begin{minipage}[b]{0.49\linewidth}
\includegraphics[width=\textwidth]{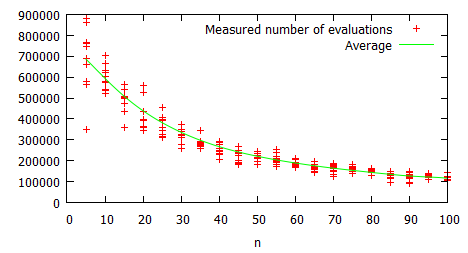}
\end{minipage}
\caption{The figure on the left shows the number of Monte Carlo trials that can be performed in a second given a spherical Pareto approximation set consisting of $n$ points. The figure on the right plots the same data as a graph of the time required for 100.000 Monte Carlo iterations, compared to the time needed for the fastest non-Monte Carlo scheme.}
\label{accuracythings}
\end{center}
\end{figure}

\begin{figure}[h!]
\begin{center}
\includegraphics[width=0.8\textwidth]{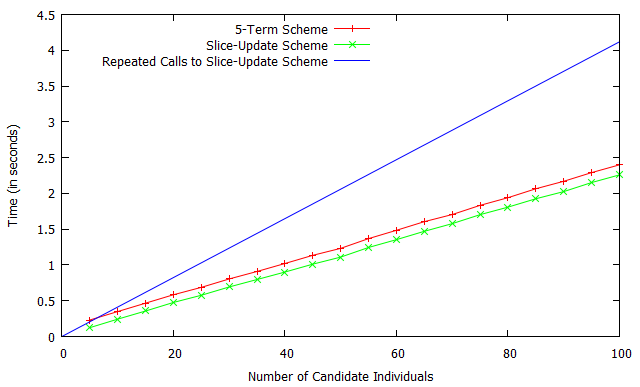}
\caption{Graph showing the time needed to simultaneously calculate the EHVI on a number of candidate points using either the 5-term or slice-update schemes, for a population size of 30. The expected time taken when simply calling the slice-update scheme on each candidate individual separately is also plotted in this graph for purposes of comparison.}
\label{multipoints}
\end{center}
\end{figure}

\clearpage

The Monte Carlo scheme is a special case, in that the time it takes to run depends on the desired accuracy, and this accuracy in turn also depends on the variance of the predictive distribution. When this variance is higher, the accuracy will be lower. For a rough idea of its performance relative to the exact calculation schemes, see Figure \ref{accuracythings}, which shows the number of Monte Carlo trials which can be performed if the algorithm is allowed to run for a second. Because of the $O(n \log n)$ time complexity of each individual trial, it is less affected by $n$ than any of the exact calculation schemes. If $n$ is large enough and the desired accuracy is low enough, it might be the faster option. However, when $n$ is reasonably small, there is no advantage to using it.

The complexity of calculating the expected hypervolume improvement of multiple points by repeatedly using one of the described algorithms is of course linear in the number of candidate individuals. Here, the 8-term, 5-term and slice-update schemes have an advantage not shared by the Monte Carlo and 2-term schemes, in that their hypervolume calculations are independent of the probability distribution for which the EHVI is being calculated. This makes it possible to calculate several expected hypervolume improvements on the same population with a relatively small corresponding increase in calculation time, because the additional calculations have complexity $O(n^3)$. It is expected to be less impressive for the slice-update scheme, as this already has a time complexity of $O(n^3)$, but the amount of overhead that is avoided might still be noticeable. To determine the impact of this advantage on the relative performance of the schemes, Figure \ref{multipoints} shows the results of using the schemes to calculate the EHVI for a vector of probability distributions instead of just one.

As can be seen, the time taken increases linearly in the number of individuals evaluated at the same time, but the constant added on top of that is larger for the 5-term scheme than for the slice-update scheme. When $n$ is 30 it is only a difference equivalent to evaluating a few more candidate individuals, however. Because the 5-term scheme is somewhat easier to implement, it might be preferable to use it if the number of candidate individuals is expected to be high in comparison to $n$.

\section{Conclusion and Future Work}
\label{thefuture}
The main results realized in this thesis are as follows: A fast algorithm for calculating the EHVI in two dimensions was proposed with runtime complexity in $O(n^2)$ (previously: $O(n^3 \log n)$).
An empirical test shows improved speed even for relatively small $n$ ($\approx 20$).
An exact calculation algorithm for calculating the EHVI in more than two dimensions is provided.
This generic algorithm has been detailed and improved in efficiency for the important tri-objective case.
It has a cubic runtime complexity in $O(n^3)$, and a further efficiency gain can be obtained by batch evaluation, i.e. re-using the data structures
for multiple EHVI computations. The algorithm are based on linear data structures and there are no large hidden constants. For three dimensions it is now possible to perform more than a hundred EHVI in 2.5 seconds for an approximation
set size of 30.
Implementations of all algorithms are made available [TODO: url] and have been validated with results from Monte Carlo algorithms.

The results open up new possibilities to construct expected improvement algorithms for multiobjective optimization, for instance by using the fast EHVI evaluation as an
infill criterion in the efficient global optimization algorithm (EGO).
Moreover, as the new exact EHVI computation methods have the same or better runtime performance compared to the Monte Carlo algorithms used so far, they can now replace these inaccurate methods.

As a side result a relationship between the expected improvement and the center of (probability) mass over single cells was established, which might in the future shed some light on the relation to alternative expected improvement formulations [Keane04] and be useful for establishing theoretical results.

\paragraph{Source code and acknowledgement}
This work is based on the honors master's thesis of Iris Hupkens \cite{Hupkens13} under the supervision of M. Emmerich and A. Deutz. The sourcecode of all algorithms in C$++$ is made available on \url{http://natcomp.liacs.nl/index.php?page=code}.


\begin{thebibliography}{}

\bibitem{Hupkens13} Iris Hupkens: Complexity Reduction and Validation of Computing the Expected
Hypervolume Improvement, Master's Thesis (with honors) published as LIACS,
Internal Report Nr. 2013-12,
August, 2013 \url{http://www.liacs.nl/assets/Masterscripties/2013-12IHupkens.pdf}

\bibitem{Shir07} Shir, O. M., Emmerich, M., B{\"a}ck, T., and Vrakking, M. J. (2007, September). The application of evolutionary multi-criteria optimization to dynamic molecular alignment. In Evolutionary Computation, 2007. CEC 2007. IEEE Congress on (pp. 4108-4115). IEEE.
\bibitem{zaefferer} Zaefferer, M., Bartz-Beielstein, T., Naujoks, B., Wagner, T., and Emmerich, M. (2013, January). A Case Study on Multi-Criteria Optimization of an Event Detection Software under Limited Budgets. In Evolutionary Multi-Criterion Optimization (pp. 756-770). Springer Berlin Heidelberg.
\bibitem{Shimoyama12}Shimoyama, K., Sato, K., Jeong, S., and Obayashi, S. (2012, June). Comparison of the criteria for updating Kriging response surface models in multi-objective optimization. In Evolutionary Computation (CEC), 2012 IEEE Congress on (pp. 1-8). IEEE.
\bibitem{Shimoyama13} Shimoyama, K., Jeong, S., and Obayashi, S. (2013, June). Kriging-surrogate-based optimization considering expected hypervolume improvement in non-constrained many-objective test problems. In Evolutionary Computation (CEC), 2013 IEEE Congress on (pp. 658-665). IEEE.
\bibitem{Couckuyt13} Couckuyt, Ivo, Dirk Deschrijver, and Tom Dhaene. "Fast calculation of multiobjective probability of improvement and expected improvement criteria for Pareto optimization." Journal of Global Optimization (2013): 1-20.
\bibitem{hypermonotonicity}
Wagner, T.; Emmerich, M.; Deutz, A. and Ponweiser, W. (2010) ``On expected-improvement criteria for model-based multi-objective optimization", in `Proc. of PPSN XI Vol. 1', Springer-Verlag, Berlin, Heidelberg, pp. 718-727.
\bibitem{hypervolume}
Fleischer, M. (2003) ``The Measure of Pareto Optima Applications to Multi-objective Metaheuristics''. Evolutionary Multi-Criterion Optimization. Second International Conference, EMO 2003, pg. 519-533.
\bibitem{exacthvi}
Emmerich, M. T M; Deutz, A.H.; Klinkenberg, J.W. (2011) ``Hypervolume-based expected improvement: Monotonicity properties and exact computation," 2011 IEEE Congress on Evolutionary Computation (CEC), pp.2147-2154

\bibitem{2dhypercomplexity}
Nicola Beume, Carlos M. Fonseca, Manuel Lopez-Ibanez, Luis Paquete, and Jan
Vahrenhold. (2009) ``On the complexity of computing the hypervolume indicator." IEEE
Trans. Evolutionary Computation, 13(5) pp. 1075-1082.

\bibitem{sachs1989}
Sacks, J., Welch, W. J., Mitchell, T. J., and Wynn, H. P. (1989) ``Design and analysis of computer experiments''. Statistical science, 4(4), 409-423.

\bibitem{mockusei}
Mockus, J., Tiesis, V., Zilinskas, A. (1978) ``The application of Bayesian methods for seeking the extremum''. In: Dixon, L., Szego, G. (Eds.), Towards Global Optimization, vol. 2. North Holland, New York, pp. 117–129.

\bibitem{jonesei}
Donald R. Jones, Matthias Schonlau, and William J. Welch. (1998) ``Efficient Global Optimization of Expensive Black-Box Functions''. J. of Global Optimization 13, 4 (December 1998), 455-492.

\bibitem{eiconvergence}
Emmanuel Vazquez and Julien Bect. (2010) ``Convergence properties
of the expected improvement algorithm with fixed mean and
covariance functions''. Journal of Statistical Planning and
Inference 140, pp. 3088-3095

\bibitem{chebyscale}
Knowles, J. (2006) ``ParEGO: A hybrid algorithm with on-line landscape approximation for expensive multiobjective optimization problems''. IEEE Transactions on Evolutionary Computation. 10 (1): 50-66.

\bibitem{keaneei}
Keane, A.J. (2006) ``Statistical improvement criteria for use in multiobjective design optimisation''. AIAA Journal, 44, (4), 879-891.

\bibitem{smsego}
Wolfgang Ponweiser, Tobias Wagner, Dirk Biermann, and Markus Vincze. (2008) ``Multiobjective Optimization on a Limited Budget of Evaluations Using Model-Assisted $\mathcal{S}$-Metric Selection''. In Proceedings of the 10th international conference on Parallel Problem Solving from Nature: PPSN X. Springer-Verlag, Berlin, Heidelberg, 784-794.

\bibitem{nsga2}
Kalyanmoy Deb, Amrit Pratap, Sameer Agarwal, and T. Meyarivan. (2000) ``A fast and elitist multi-objective genetic
algorithm: NSGA-II''.

\bibitem{spea2}
 E. Zitzler, M. Laumanns and L. Thiele. (2001) ``SPEA2: Improving the Strength Pareto Evolutionary Algorithm''.

\bibitem{smsemoa}
Michael Emmerich, Nicola Beume, and Boris Naujoks. (2005) ``An EMO Algorithm Using the Hypervolume Measure as Selection Criterion''. In 2005 Intl Conference, March 2005, pages 62-76.

\bibitem{covarmat}
Christian Igel, Nikolaus Hansen, and Stefan Roth. (2007) ``Covariance Matrix Adaptation for Multi-objective Optimization''. Evol. Comput. 15, 1 (March 2007), 1-28.

\bibitem{kriging}
Williams, Christopher K.I. (1998) ``Prediction with Gaussian processes: From linear regression to linear prediction and beyond". In M. I. Jordan. Learning in graphical models. MIT Press. pp. 599–612.

\bibitem{fubini}
Fubini, G. "Sugli integrali multipli." (1958) Opere scelte, Vol. 2. Cremonese, pp. 243-249.

\bibitem{twister}
Matsumoto, M.; Nishimura, T. (1998) ``Mersenne twister: a 623-dimensionally equidistributed uniform pseudo-random number generator". ACM Transactions on Modeling and Computer Simulation 8 (1): 3–30

\bibitem{boxmuller}
G. E. P. Box, Mervin E. Muller. (1958) A Note on the Generation of Random Normal Deviates. The Annals of Mathematical Statistics, Vol. 29, No. 2. pp. 610-611

\bibitem{emmerichphd} Emmerich, M. (2005). Single-and multi-objective evolutionary design optimization assisted by gaussian random field metamodels. Dissertation, TU Dortmund, Informatik, Eldorado, http://hdl.handle.net/2003/21807.
\bibitem{kumano} Kumano, T., Jeong, S., Obayashi, S., Ito, Y., Hatanaka, K., and Morino, H. (2006). Multidisciplinary design optimization of wing shape with nacelle and pylon. In European Conference on Computational Fluid Dynamics ECCOMAS CFD.
\bibitem{miettinen} Miettinen, K. (1999). Nonlinear Multiobjective Optimization, volume 12 of International Series in Operations Research and Management Science.

\bibitem{zitzler} Zitzler, E.,  Thiele, L. (1998, January). Multiobjective optimization using evolutionary algorithms—a comparative case study. In Parallel problem solving from nature—PPSN V (pp. 292-301). Springer Berlin Heidelberg
\bibitem{viviane} Zitzler, E., Thiele, L., Laumanns, M., Fonseca, C. M., and Da Fonseca, V. G. (2003). Performance assessment of multiobjective optimizers: An analysis and review. Evolutionary Computation, IEEE Transactions on, 7(2), 117-132.
\bibitem{fleischer} Fleischer, M. (2003, January). The measure of Pareto optima applications to multi-objective metaheuristics. In Evolutionary multi-criterion optimization (pp. 519-533). Springer Berlin Heidelberg.
\bibitem{auger} Auger, A., Bader, J., Brockhoff, D., and Zitzler, E. (2009, January). Theory of the hypervolume indicator: optimal $\mu$-distributions and the choice of the reference point. In Proceedings of the tenth ACM SIGEVO workshop on Foundations of genetic algorithms (pp. 87-102). ACM.
\bibitem{bringmann} Bringmann, K., and Friedrich, T. (2010, July). The maximum hypervolume set yields near-optimal approximation. In Proceedings of the 12th annual conference on Genetic and evolutionary computation (pp. 511-518). ACM.
\bibitem{Wollk}Laniewski-Wollk, P., Obayashi S.,  Jeong, S. (2010), Development of expected improvement for multi-objective problems, in Proceedings of 42nd Fluid Dynamics Conference/Aerospace Numerical, Simulation Symposium (CD ROM), June 2010
\bibitem{Koch} Koch, P. (2013). Efficient tuning in supervised machine learning (Doctoral dissertation, Leiden Institute of Advanced Computer Science (LIACS), Faculty of Science, Leiden University).



\end{thebibliography}
\end{document}